\documentclass[aps,preprint,pdftex,superscriptaddress]{revtex4}
\usepackage{amssymb}
\usepackage{amsmath}
\usepackage{graphicx}
\usepackage{bm}
\usepackage{color}
\usepackage{subfigure}
\usepackage{epstopdf}

\def\Bbf{\mathbf B}

\def\Jbf{\mathbf J}

\def\kbf{\mathbf k}

\def\ubf{\mathbf u}

\begin{document}

\title{Plasma flow evolution in response to resonant magnetic perturbation in a tokamak}

\author{Xingting Yan}
\affiliation{CAS Key Laboratory of Geospace Environment and Department of Engineering and Applied Physics, University of Science and Technology of China, Hefei, Anhui 230026, China}

\author{Ping Zhu}
\email[E-mail:~]{zhup@hust.edu.cn}
\affiliation{International Joint Research Laboratory of Magnetic Confinement Fusion and Plasma Physics, State Key Laboratory of Advanced Electromagnetic Engineering and Technology, School of Electrical and Electronic Engineering, Huazhong University of Science and Technology, Wuhan, Hubei 430074, China}
\affiliation{Department of Engineering Physics, University of Wisconsin-Madison, Madison, Wisconsin 53706, USA}

\author{Wenlong Huang}
\affiliation{School of Computer Science and Technology, Anhui University of Technology, Ma'anshan, Anhui 243002, China}

\date{\today}
\begin{abstract}
Externally applied non-axisymmetric magnetic fields such as error field and resonant magnetic perturbation (RMP) are known to influence the plasma momentum transport and flow evolution through plasma response in a tokamak, whereas the evolution of plasma response itself strongly depends on the plasma flow as well. The nonlinear interaction between the two have been captured in the conventional error field theory with a ``no-slip'' condition, which has been recently extended to allow the ``free-slip'' condition. For comparison with simulations, we solve for the nonlinear plasma response and flow evolution driven by a single-helicity RMP in a tokamak, using the full resistive MHD model in the initial-value code NIMROD. Time evolution of the parallel (to $\kbf$) flow or ``slip frequency'' profile and its asymptotic steady state obtained from the NIMROD simulations are compared with both conventional and extended nonlinear response theories. Here $\kbf$ is the wave vector of the propagating island. Good agreement with the extended theory with ``free-slip'' condition has been achieved for the parallel flow profile evolution in response to RMP in all resistive regimes, whereas the difference from the conventional theory with the ``no-slip'' condition tends to diminish as the plasma resistivity approaches zero.
\end{abstract}

\maketitle
\section{Introduction}

Plasma responses to external non-axisymmetric magnetic fields such as error field (EF) and resonant magnetic perturbation (RMP) are well known to play significant roles in many areas of tokamak physics. The intrinsic error field from the tokamak coil system can lead to the locking of rotating tearing modes in plasma~\cite{nave90a}, and the subsequent growth of locked tearing mode often gives rise to major disruptions. In general, the externally applied resonant and non-resonant magnetic perturbations can brake or accelerate toroidal rotations through plasma response, whereas toroidal rotation can effectively influence various MHD and transport processes in both core and edge plasmas (e.g.~\cite{monticello79a,lazzaro88a,lazzaro02a,coelho04a}).
In experiments on DIII-D~\cite{d3d}, it was discovered that an RMP can suppress most type-I ELMs in high confinement plasmas while leaving the transport barrier or core confinement nearly intact~\cite{d3,iter,elm,d1,d2}. In KSTAR~\cite{kstar}, ELMs are completely suppressed by applying $n=1$ non-axisymmetric magnetic perturbations~\cite{k1} where the toroidal rotation is also slowed down~\cite{k2} (here $n$ is the toroidal mode number). Recently, evidence of a nonlinear transition from mitigation to suppression of the ELM by using $n=1$ and $2$ RMPs have been observed in the EAST tokamak~\cite{sunyw16a,sunyw17a}. Due to the emerging and promising potential of RMP as an effective and versatile tool for controlling plasma properties and behaviors in tokamaks, the subject on the interactions between RMPs and plasmas has received continued interests. 

At least two key physics processes are believed to involve in the interaction between RMP and tokamak plasma, namely, the plasma response to RMP in presence of plasma flow, and the braking and acceleration of plasma flow due to the resonant and non-resonant torques induced by plasma response. Here in this paper, we refer the plasma ``flow'' or ``rotation'' to the surface-averaged plasma velocity, or equivalently, the $(0, 0)$ component of the Fourier transform of the plasma velocity field in poloidal and toroidal directions. On the one hand, plasma flow provides screening effects on the penetration of RMP field, significantly affecting the amplitude and structure of plasma response. On the other hand, plasma response produces resonant and non-resonant torques, such as the electromagnetic torque and the neoclassical toroidal viscous torque, that largely contribute to the evolution of plasma flow profile. Such an interaction is inevitably and highly nonlinear.
Due to the complexity associated with the intrinsically nonlinear nature of plasma response, the two processes have been studied theoretically and numerically on the basis of kinetic and fluid models, in both linear and nonlinear regimes, at different levels of sophistication and self-consistency.

In the linear regime, the time advance of plasma response can be described by the solutions to the Taylor problem, i.e. the forced magnetic reconnection induced by perturbation of boundary magnetic flux, based on the theory developed by~\citet{hahm85a}, where for simple geometries, analytical solutions of the linear plasma response can be obtained. The Hahm and Kulsrud (HK) theory is later extended from slab to cylindrical configurations~\cite{fitzpatrick91a}, and from static plasma to plasma in presence of equilibrium flow~\cite{huangwl20a}. Although the linear solutions from HK-type of theory by design do not take into account the effects of plasma response on the equilibrium flow itself, they do predict plasma response in both magnetic and velocity fields that are of same helicity as the external perturbation at boundary.

In the nonlinear regime, previous theory on the EF penetration has been applied to the analysis of plasma response to RMPs, where the nonlinear interaction between plasma flow and response have been modeled within quasi-linear approximation for a coupled system of torque balance and magnetic island evolution equations in the Rutherford regime~\cite{fitzpatrick93a,fitzpatrick14a}. Such a theory is able to model the plasma flow evolution driven by the electromagnetic torque induced by RMP and balanced by the viscous torque. However, the ``no-slip'' condition is often imposed in previous theory, where the magnetic island is assumed to move together with plasma flow, which is not always satisfied in simulations or experiments (e.g. Sec.~\ref{sec:com} and~\cite{yanw18a}). Recently, such a nonlinear or quasi-linear model for plasma response and flow evolution has been recently extended to allow the ``free-slip'' condition, where the island phase equation extending beyond the ``no-slip'' condition is naturally obtained along with the conventional Rutherford equation for island width growth~\cite{huangwl20a}.

Numerical calculations have also been developed for plasma response to RMPs in tokamaks since 2000s. Most of these are based on linearized MHD models, and they reach best agreement with theory in the slab configuration(e.g.~\cite{beidler18a}). For cylindrical and toroidal configurations, the comparisons between theory and numerical results are rare and less certain, even in linear calculations. For all these linear calculations of plasma response, the plasma flow, if present, is held fixed. Apparently, only in nonlinear or at least quasi-linear simulations can plasma flow evolution in response to RMP be evaluated. For example, the linear plasma response to Dynamic Ergodic Divertor (DED) on TEXTOR obtained earlier from a cylindrical single-fluid resistive reduced MHD model, is used to calculate the resonant $\Jbf\times\Bbf$ torque in quasi-linear approximation, which, along with collisional viscous torques, is further applied to solving for toroidal rotation~\cite{kikuchi06a}. Depending on the DED frequency, it is found that the external magnetic perturbation can either brake or accelerate toroidal rotation. Later, the cylindrical 4-field reduced two-fluid MHD model has extended the simulation on plasma response to RMP on DIII-D. The simulations remain quasi-linear in nature, where the perturbation harmonics can, by interacting with themselves, modify the profiles of the axially and the azimuthally symmetric (i.e. $(0,0)$) Fourier component~\cite{nardon10a}. RMP screening (penetration) occurs when the perpendicular electron flow $v_{e\perp}$ becomes finite (zero) at the rational surface, as demonstrated in the two-fluid simulations. The quasi-linear approximation is also adopted in the toroidal resistive full MHD model in the MARS-Q code for the calculation of toroidal flow damping due to plasma response in the MAST experiment, where linear plasma response is used to calculate both resonant $\Jbf\times\Bbf$ and neoclassical toroidal viscosity (NTV) torques~(e.g.~\cite{liuyq12a,liuyq13a}). Fully nonlinear simulations of plasma response to RMP along with toroidal rotation evolution on DIII-D in the ITER-like regime are performed using a cylindrical 4-field reduced MHD model including NTV torque~\cite{becoulet09a}. Screening of resonant component of response increases with stronger toroidal rotation and lower resistivity, and both toroidal rotation damping and acceleration are obtained in those simulations. Recent nonlinear resistive single-fluid MHD simulations of plasma response has been carried out in a cylindrical configuration using the NIMROD code, where only the poloidal rotation evolution is considered~\cite{akcay20a}. Many other early and recent nonlinear simulations on plasma response using various MHD models and codes have been reported, where the attentions are directed towards other key aspects of the response process instead of the flow evolution (e.g.~\cite{izzo08b},~\cite{huqm20a}). 

In this work, we perform a comparative study on the evolution of the plasma flow in response to RMP in a tokamak, using both nonlinear simulations from the full resistive single-fluid MHD model implemented in the NIMROD code for the complete toroidal geometry, and theory predictions  for nonlinear plasma response from an extended model that allows the ``free-slip'' condition for the island-flow phase relation. Time evolution of the parallel (to $\kbf$) flow or ``slip frequency'' profile and its asymptotic steady state obtained from the NIMROD simulations are compared with both the conventional and the extended theories for nonlinear plasma response. Here $\kbf$ is the wave vector of the propagating island. Good agreement with the extended theory with ``free-slip'' condition has been achieved for the parallel flow profile evolution in response to RMP in all resistive regimes, whereas the difference from the conventional theory with the ``no-slip'' condition diminishes as the plasma resistivity approaches zero.

The rest of the paper is organized as follows. In Sec.~\ref{sec:the} we introduce the extended theory model for RMP-island interaction employed in this work. In Sec.~\ref{sec:sim}, the set up for the NIMROD simulations of the nonlinear plasma response to RMP is described and explained. This is followed by Sec.~\ref{sec:com}, where the NIMROD simulation results for the parallel flow evolution are compared with numerical solutions of the extended theory for the RMP-island interaction. Finally we give summary and discussion in the Sec.~\ref{sec:sum}.

%

\section{Theory models for nonlinear plasma response}
\label{sec:the}
The key physics of nonlinear plasma response and flow evolution induced by RMP has often been described in theory by a coupled system of equations that govern the magnetic island growth and the rotation torque balance. One such theory derives from the model for tearing mode locking due to error field, where are believed to share the similar physics and equations as the plasma response and RMP, respectively.  In particular, a slightly generalized error field model for nonlinear plasma response consists of the following system of equations set in a cylindrical tokamak, which include the torque balance equations for both toroidal and poloidal rotations~\cite{fitzpatrick93a,fitzpatrick00a,fitzpatrick02a} (See also, for example, Appendix~\ref{app:tor_pol}):
\begin{equation}
r\rho\frac{\partial\Delta\Omega_\phi}{\partial t}-\frac{\partial}{\partial r}\left(r\mu\frac{\partial\Delta\Omega_\phi}{\partial r}\right)=\frac{\hat{T}_z}{4\pi^2R_0^3}\delta(r-r_s),
\label{eq:tor_rot}
\end{equation}
\begin{equation}
\frac{\partial\Delta\Omega_\phi(0,t)}{\partial r}=\Delta\Omega_\phi(a,t)=0,
\end{equation}
\begin{equation}
r^3\rho\frac{\partial\Delta\Omega_\theta}{\partial t}-\frac{\partial}{\partial r}\left(r^3\mu\frac{\partial\Delta\Omega_\theta}{\partial r}\right)=-\frac{m\hat{T_z}}{n4\pi^2R_0}\delta(r-r_s),
\label{eq:pol_rot}
\end{equation}
\begin{equation}
\frac{\partial\Delta\Omega_\theta(0,t)}{\partial r}=\Delta\Omega_\theta(a,t)=0,
\end{equation}
where the toroidal (poloidal) rotation frequency $\Omega_\phi=\Omega_{\phi 0}+\Delta\Omega_\phi$ ($\Omega_\theta=\Omega_{\theta 0}+\Delta\Omega_\theta$), and $\Omega_{\phi 0}$ ($\Omega_{\theta 0}$) is the initial equilibrium toroidal (poloidal) rotation frequency. The modified Rutherford equation for the magnetic island width growth from plasma response~\cite{rutherford73a,arcis06a}:
\begin{equation}
\frac{\tau_R}{1.22r_s^2}\frac{dW}{dt}=\Delta_l'+\Delta_{nl}'+\Delta_c'(\frac{W_c}{W})^2\cos\varphi,
\label{eq:width}
\end{equation}
and the no-slip condition for the island phase variation~\cite{fitzpatrick93a}
\begin{equation}
\frac{d\varphi}{dt}=\omega_s=-\mathbf{k}\cdot\mathbf{u}_0=n\Omega_{\phi s}-m\Omega_{\theta s}.
\label{eq:phase_nslip}
\end{equation}
Here, $\theta$ ($\phi$) is the poloidal (toroidal) angle, $\Omega_\theta$ ($\Omega_\phi$) the poloidal (toroidal) rotation frequency of the flux surface-averaged plasma flow $\mathbf{u}_0$, $m$ ($n$) the poloidal (toroidal) Fourier mode number as defined in the Fourier harmonic component $\exp{[i(m\theta-n\phi)]}$, $\hat{T}_z$ the toroidal electromagnetic torque induced by RMP at rational surface denoted as $r_s$, $W$ the island width from the resonant magnetic response at $r_s$, $W_c$ the equivalent island width for the external resonant magnetic perturbation, and $\varphi$ the phase difference between the resonant magnetic response at $r_s$ and the external RMP at boundary. In the modified Rutherford equation~(\ref{eq:width}), $\tau_R$ is the resistive time, $\Delta_l'$ , $\Delta_{nl}'$ and $\Delta_c'$ represent linear driver, nonlinear saturation, and external RMP effects respectively, which are functions of the island width $W$ and relative phase $\varphi$. The details of these function dependence and the corresponding definitions can be found in Ref.~\cite{huangwl15a} and thus are not repeated here.

Recently, the above system of equations~(\ref{eq:tor_rot}) to~(\ref{eq:phase_nslip}) have been re-derived purely from a 2-field reduced MHD equations in absence of toroidal rotation, which have not only recovered most terms in Eqs.~(\ref{eq:tor_rot}) to~(\ref{eq:phase_nslip}) but also found a natural extension to the island phase equation that allows the more general ``free-slip'' condition for the phase relation between plasma response and its corresponding external RMP as in the following equation~\cite{huangwl20a}:
\begin{equation}
\frac{d\varphi}{dt} = \omega_s - \frac{\sqrt2 a^2}{2A\tau_R}\Delta_c'\frac{W_c^2}{W^3}sin\varphi,
\label{eq:phase_fslip}
\end{equation}
where $\omega_s=-\mathbf{k}\cdot\mathbf{u}_0$ on rational surface, and $A\simeq 0.7$.  The origin of the ``free-slip'' term in Eq.~(\ref{eq:phase_fslip}) comes from the resistivity within the resistive layer and the consequent breaking of frozen-in condition there. Thus the appearance of the ``free-slip'' term in the island phase equation in (\ref{eq:phase_fslip}) is a natural extension to the conventional Rutherford equation, where the island phase is simply assumed a constant or ignored~\cite{rutherford73a}. For comparisons with three-dimensional (3D) simulation results that involve both toroidal and poloidal rotations, we further extend the theory model by keeping the toroidal torque balance equation (\ref{eq:tor_rot}) and including the toroidal component in the phase equation (\ref{eq:phase_fslip}), i.e. $\omega_s=-\mathbf{k}\cdot\mathbf{u}_0=n\Omega_{\phi s}-m\Omega_{\theta s}$. For the sake of discussions hereafter, we refer to Eqs. (\ref{eq:tor_rot})-(\ref{eq:phase_nslip}) as the ``no-slip'' or ``NS'' plasma response model, and Eqs. (\ref{eq:tor_rot})-(\ref{eq:width}) along with Eq. (\ref{eq:phase_fslip}) as the ``free-slip'' or ``FS'' plasma response model. Numerical solutions to both ``NS'' and ``FS'' models are then subjects to comparison with simulation results on the time evolution of slip frequency $\mathbf{k}\cdot\mathbf{u}_0$. We later refer to these numerical solutions as the ``Newcomb'' solutions, since these quasi-linear model are build upon the linear solutions of the corresponding Newcomb equations~\cite{huangwl15a,huangwl16a}. We report the simulation-theory comparison results in Sec.~\ref{sec:com}.


\section{Simulation model and setup}
\label{sec:sim}
The simulations of plasma response to RMP in a tokamak are based on the full single-fluid resistive MHD model implemented in the NIMROD code~\cite{sovinec04a} 
\begin{eqnarray}
\frac{d\rho}{dt} &=& -\rho\nabla\cdot\ubf + D\nabla^2\rho 
\label{eq:mhd_den} \\
\rho\frac{d\ubf}{dt} &=& -\nabla p
+\mathbf{J}\times\mathbf{B}
-\nabla\cdot\rho\nu\nabla\mathbf{u}
\\
\frac{N}{\gamma-1}\frac{dT}{dt}&=&-\frac{p}{2}\nabla\cdot\mathbf{u}-\nabla\cdot\mathbf{q} \\
\frac{\partial\Bbf}{\partial t} &=& -\nabla\times\mathbf{E} \\
\mathbf{E}&=&-\ubf\times\Bbf+\eta\mathbf{J} \\
\mu_0\mathbf{J}&=&\nabla\times\mathbf{B}
\label{eq:mhd_b}
\end{eqnarray}
where $d/dt=\partial/\partial t+\ubf\cdot\nabla$, $\gamma$ is the adiabatic index, $\rho$ ($N$) the mass (number) density, $\mathbf{u}$ the plasma velocity, $p$ the total pressure, $D$ the mass diffusivity,  $\nu$ the kinematic viscosity, $\mathbf{q}=-N[\kappa_\parallel\mathbf{bb}+\kappa_\perp(\mathbf{I}-\mathbf{bb})]\cdot\nabla T$, with $\mathbf{b}=\mathbf{B}/|\mathbf{B}|$ being the local magnetic direction unit vector, $\kappa_\parallel$ ($\kappa_\perp$) the parallel (perpendicular) thermal conductivity with respect to the local magnetic field direction, $\eta$ the resistivity, and the rest of the symbols are conventional.

\begin{figure}[htbp]
\centering
\subfigure{\includegraphics[width=0.75\textwidth,angle=0]{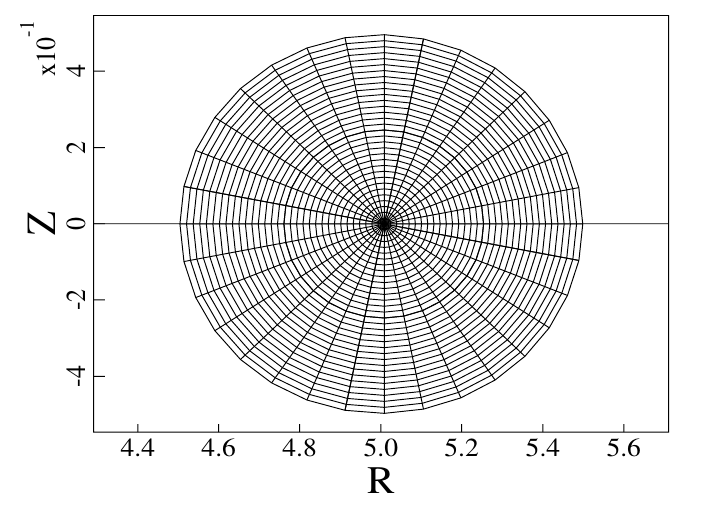}}

\subfigure{\includegraphics[width=0.75\textwidth,angle=0]{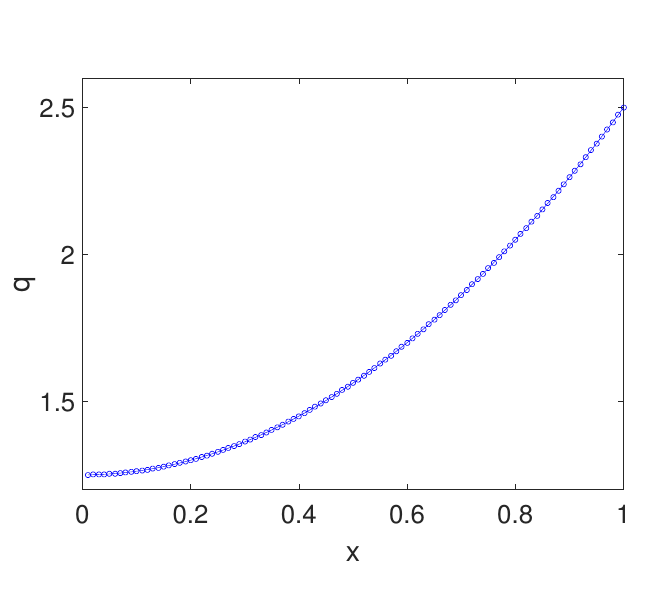}}
\caption{Finite element mesh used in NIMROD simulations aligned with the flux surfaces (upper) and the corresponding $q$ profile of a circular-shaped limiter tokamak equilibrium (lower).}
\label{fig:equ_q_prof}
\end{figure}

A model equilibrium for the limiter tokamak with a circular-shaped boundary has been obtained from the ESC code~\cite{zakharovl99a} and used in this study. The pressure profile is assumed uniform, and the safety factor profile has the form of $q(x)=1.25(1+x^2)$, where $x=\sqrt{\psi/\psi_a}$ is the normalized minor radius defined with the poloidal flux function $\psi$ and its value $\psi_a$ at tokamak boundary (Fig.~\ref{fig:equ_q_prof}).

NIMROD simulations are set up for calculating nonlinear plasma response to the RMP that is prescribed as a fixed boundary condition at tokamak wall location. We consider a static RMP with its normal component $B_\psi(\theta,\phi)=B_{\psi a}\cos(m\theta-n\phi)$ prescribed at the circular-shaped boundary of a model limiter tokamak, where the minor radius $a=0.5 m$ and the major radius $R_0=5 m$. Here, all quantities are in SI units unless otherwise noted. The uniform equilibrium pressure considered is in a low plasma $\beta$ regime, with $\beta=\mu_0p_0/B_0^2=0.0045$, where $p_0$ and $B_0$ are the equilibrium values of pressure and magnetic field magnitude at magnetic axis respectively. The simulations are initialized with a non-uniform toroidal rotation as a part of the axisymmetric equilibrium fields. A single helicity RMP is considered with $m/n = 2/1$. Eqs. (\ref{eq:mhd_den}) to (\ref{eq:mhd_b}) are numerically solved and advanced using the NIMROD code to calculate the nonlinear plasma response to the prescribed RMP boundary condition, and the resulting profile evolution of the slip frequency $\mathbf{k}\cdot\mathbf{u}_0$ from simulations are compared with the numerical solutions from both NS (``no-slip'') and FS (``free-slip'') theory models for nonlinear plasma response (see Sec.~\ref{sec:the}) for several relevant parameter regimes in next section.

\section{Parallel flow evolution: simulation and theory results}
\label{sec:com}
For our comparison study, the tokamak equilibrium is initialized with a non-uniform toroidal rotation before the application of RMP. The toroidal rotation frequency is a function of minor radius as in $\Omega_{\phi 0}=\Omega_0(1-x^5)$ (i.e. $\Omega_{\theta 0}=0$ is assumed at $t=0$). Once the RMP is turned on, the toroidal rotation profile would evolve along with other parts of the plasma response. For a static RMP, and for those cases of response from tokamak plasma, including the plasma flow, that eventually reach a steady state, there are two different categories of states, where the plasma parallel flow $\mathbf{k}\cdot\mathbf{u}_0$ either drops to zero or remains finite value on the resonant flux surface, which we refer to as the``locked state'' and the ``unlocked state'' of plasma flow, respectively. We report our findings on the comparison study for each of the two types of response states in this section.

In all the nonlinear NIMROD simulations of plasma response presented in this paper, the toroidal Fourier modes with mode number $n=0-1$ are included, which are found numerically convergent with respect to the numbers of toroidal Fourier components. For both the NS and the FS theory models described in Sec.~\ref{sec:the}, the coupled nonlinear system of equations (\ref{eq:tor_rot}) to (\ref{eq:phase_nslip}) or (\ref{eq:phase_fslip}) are decomposed first in Bessel function space and then solved numerically for each Bessel component. We compare primarily the $\mathbf{k}\cdot\mathbf{u}_0$ profile evolution in response to RMP from both NIMROD simulations and the numerical solutions of theory models.

\subsection{Locked state of plasma flow}
\label{subsec:locked}
\begin{figure}[htbp]
\centering
\subfigure{\includegraphics[width=0.49\textwidth,angle=0]{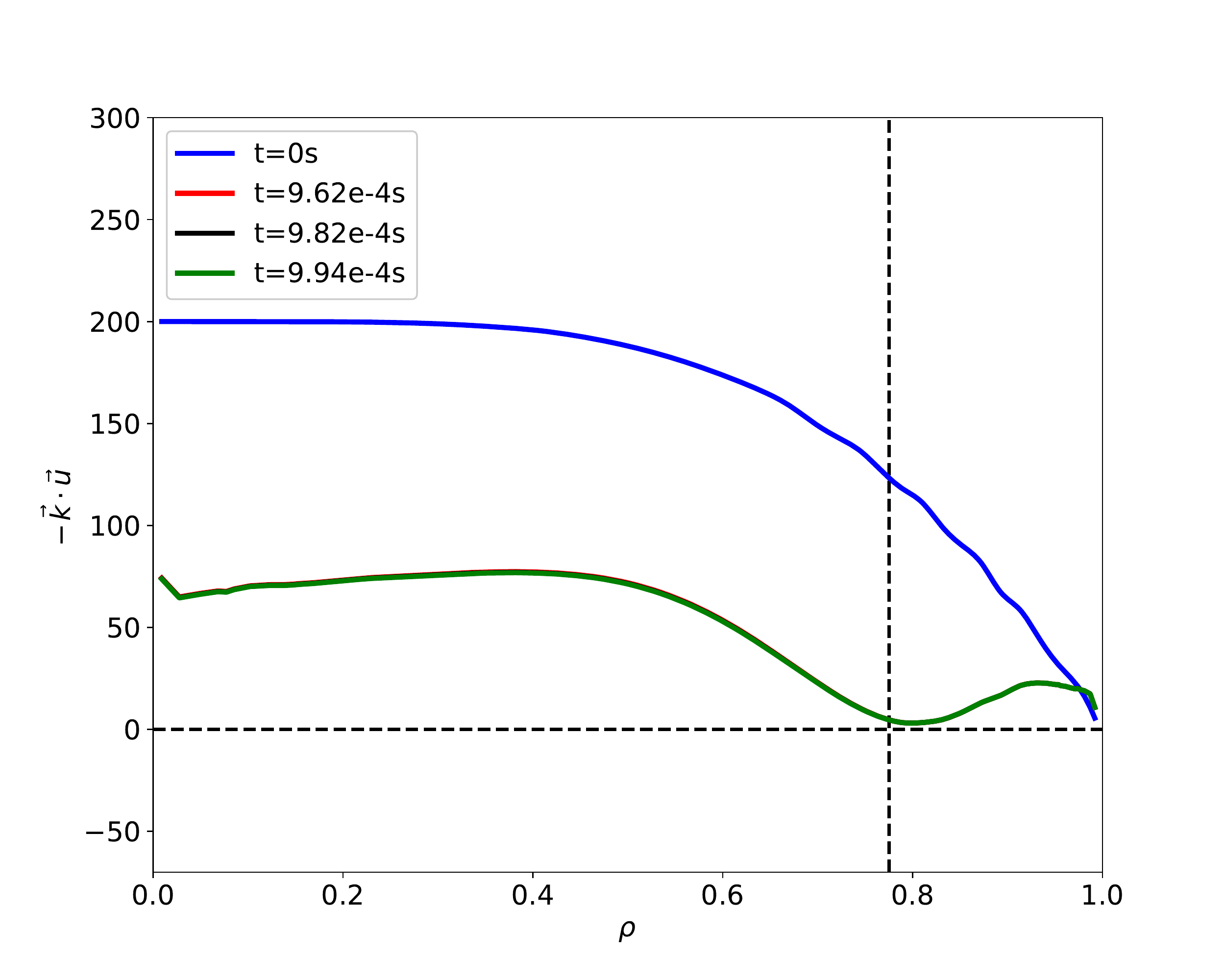}}
\subfigure{\includegraphics[width=0.49\textwidth,angle=0]{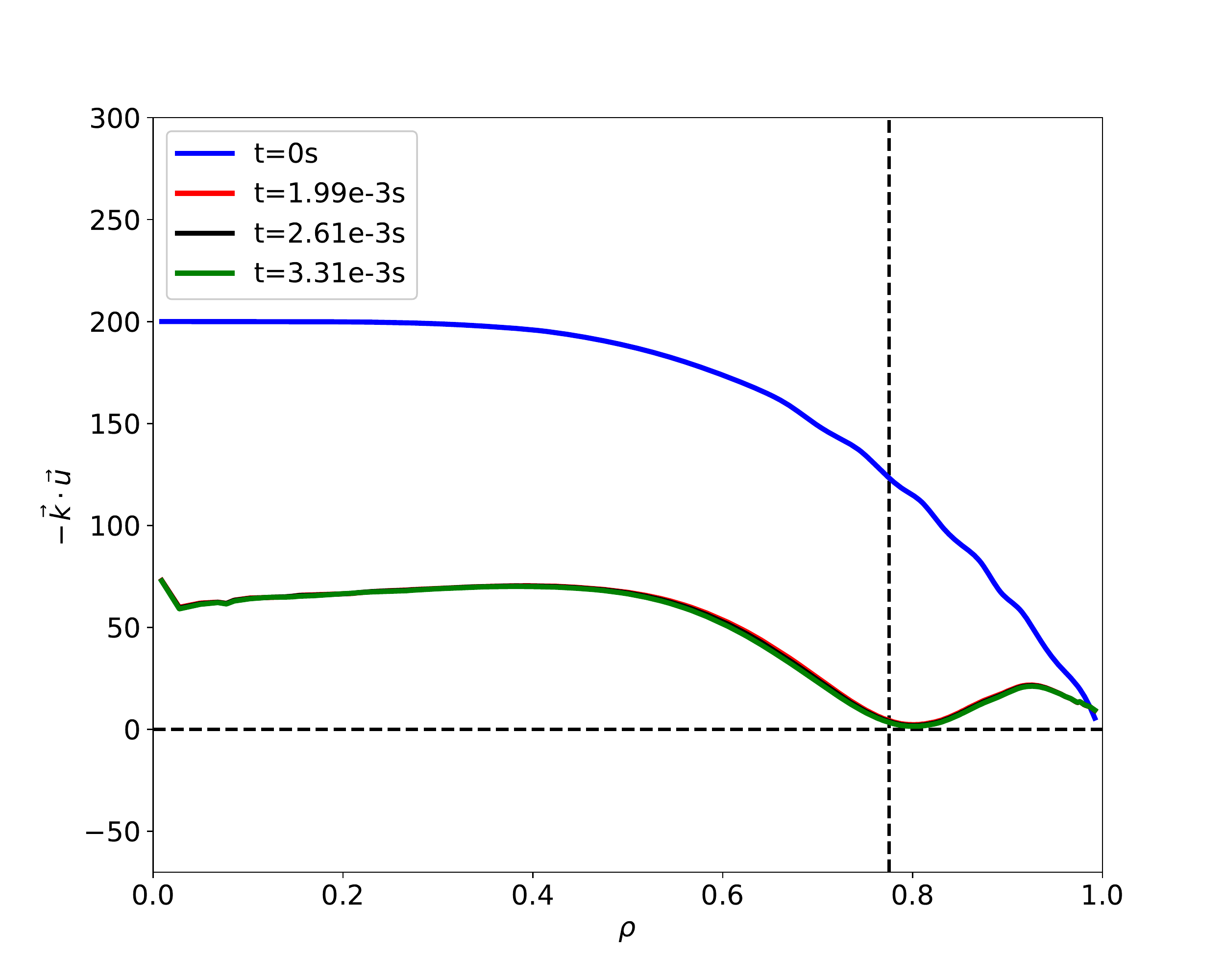}}

\subfigure{\includegraphics[width=0.49\textwidth,angle=0]{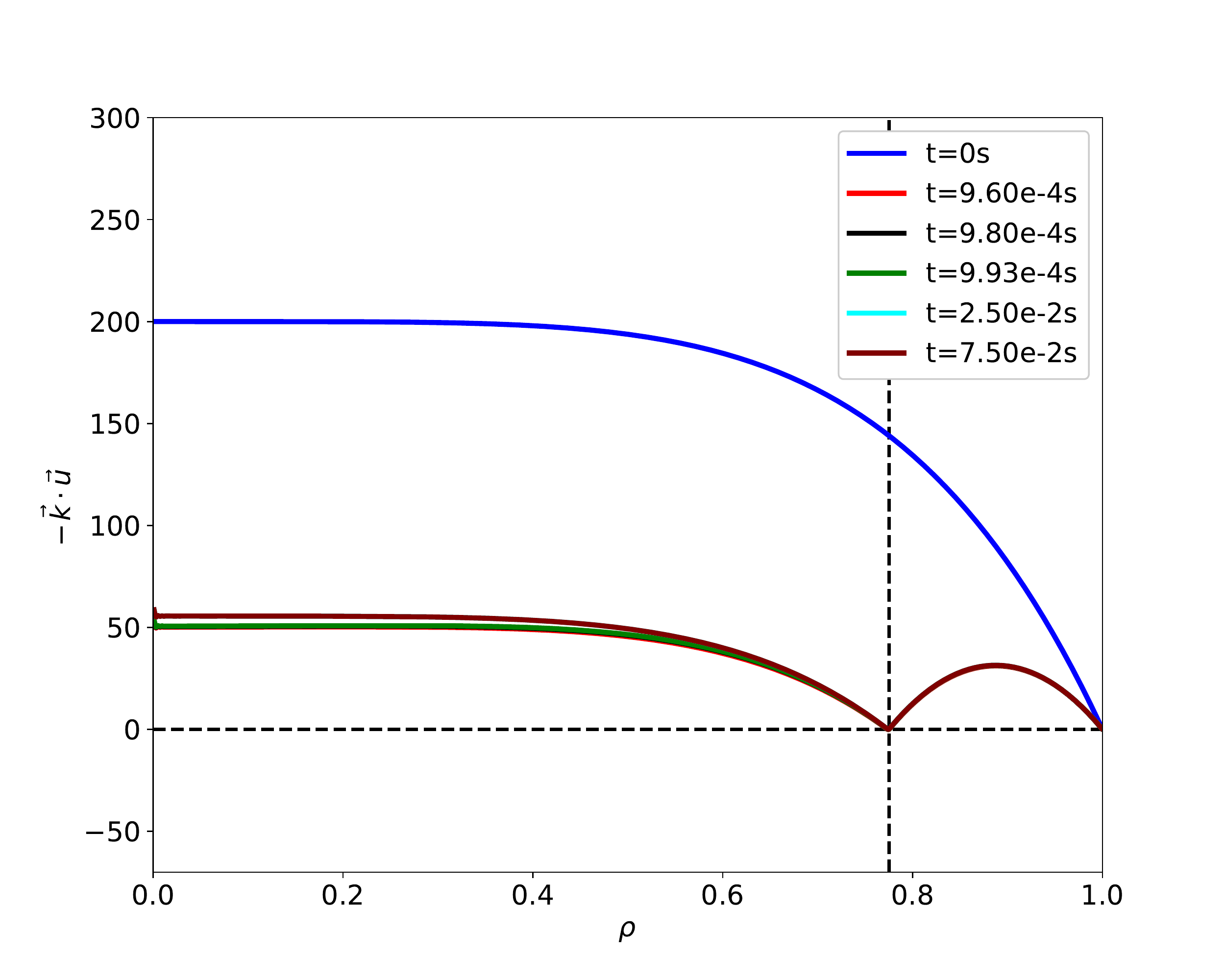}}
\subfigure{\includegraphics[width=0.49\textwidth,angle=0]{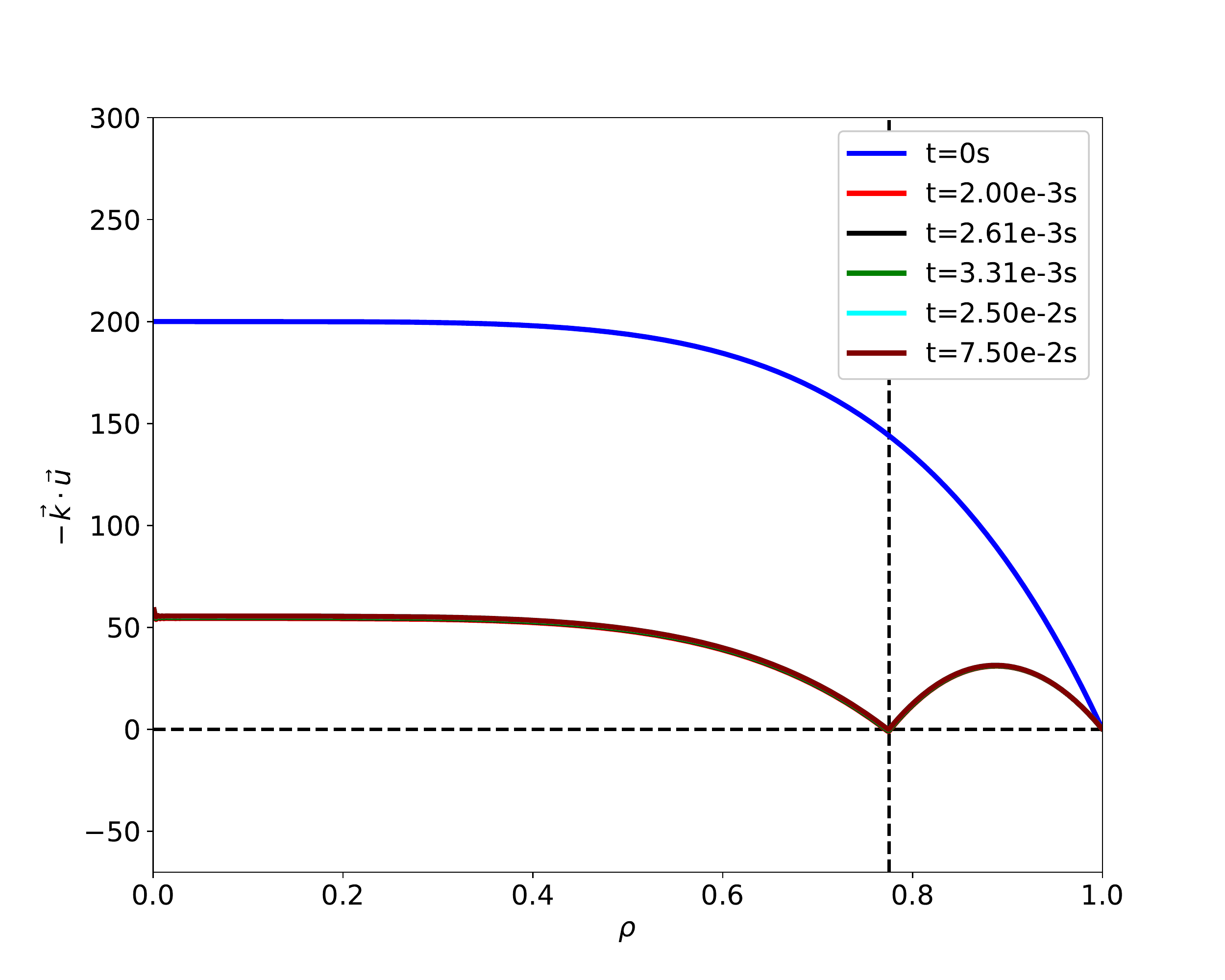}}

\subfigure{\includegraphics[width=0.49\textwidth,angle=0]{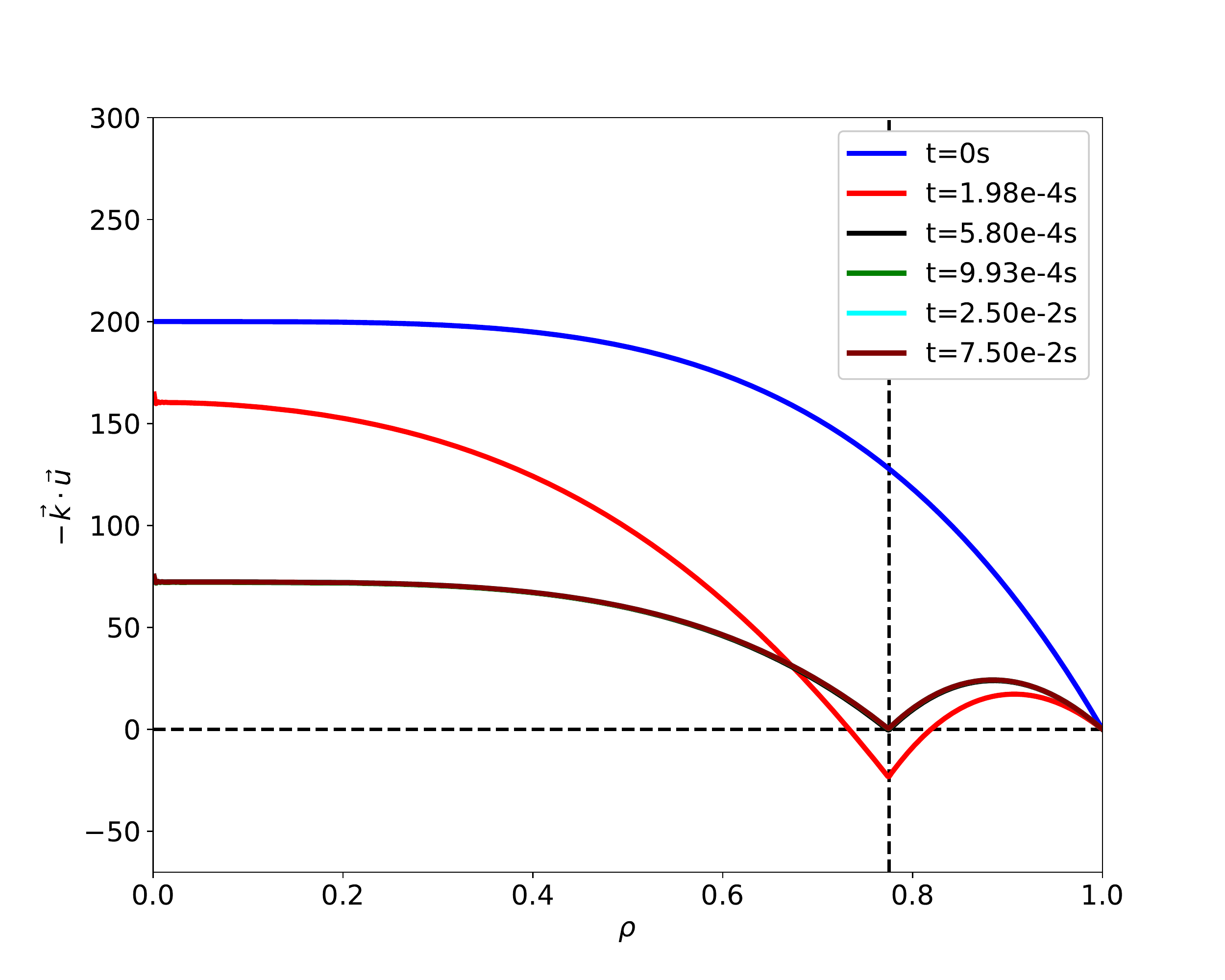}}  
\subfigure{\includegraphics[width=0.49\textwidth,angle=0]{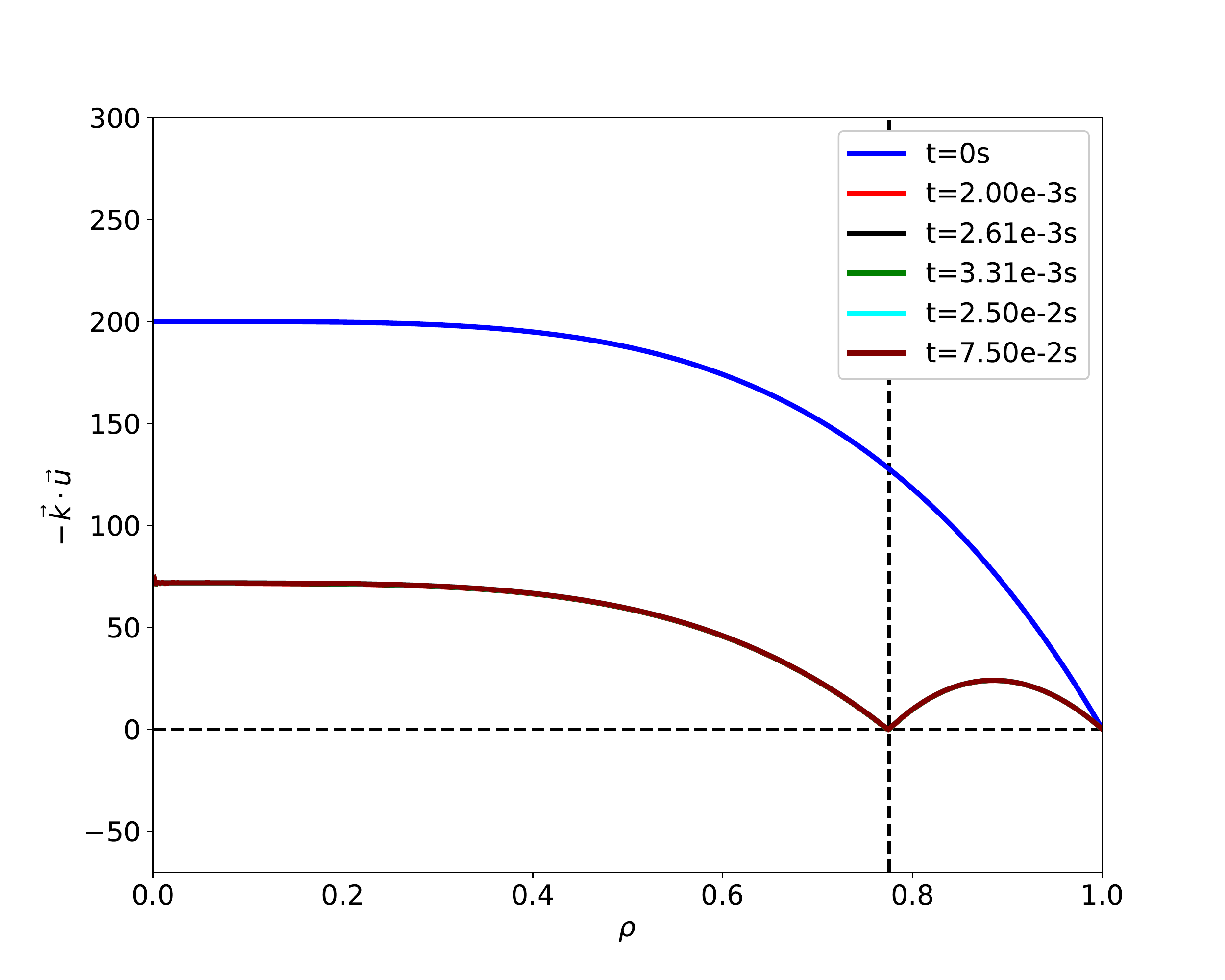}}  
\caption{Radial profiles of $\mathbf{k}\cdot\mathbf{u}_0$ at different time slices from NIMROD simulations (upper), and Newcomb solutions to the NS theory model (middle) and the FS theory model (lower) for the $S=3\times10^5$, $Pr_m=40$ (left) and the $S=3\times10^6$, $Pr_m=400$ cases (right).}
\label{fig:locked_ku_profile}
\end{figure}

For a given RMP with its helicity and phase prescribed in Sec.~\ref{sec:sim}, whether the plasma flow can reach the ``locked state'' is determined largely by the amplitude of RMP, the magnitude of the initial plasma flow, and the plasma viscosity value (See, for example, Appendix~\ref{app:w_wc}). For the initial toroidal rotation profile specified above, we first consider two representative ``locked state'' cases of plasma response for comparison study where for both cases the core toroidal rotation frequency $\Omega_0=2\times10^2 rad/s$, the uniform number density $N=10^{18}m^{-3}$, and the equivalent island width of the RMP amplitude $W_C/a=0.292$. The two cases differ only in the Lundquist number and the magnetic Prandtl number, which are $S=3\times10^5$, $Pr_m=40$ in one case, and $S=3\times10^6$, $Pr_m=400$ in another. The $\mathbf{k}\cdot\mathbf{u}_0$ profile evolution are extracted from the nonlinear plasma response results at several representative time slices for comparisons, which show good agreement in timing and radial profile between the NIMROD results (Fig.~\ref{fig:locked_ku_profile}, upper panels) and the theory results from NS model (Fig.~\ref{fig:locked_ku_profile}, middle panels) and FS model (Fig.~\ref{fig:locked_ku_profile}, lower panels), for both the $S=3\times10^5$, $Pr_m=40$ case (Fig.~\ref{fig:locked_ku_profile}, left column) and the $S=3\times10^6$, $Pr_m=400$ case (Fig.~\ref{fig:locked_ku_profile}, right column). The simulation and theory calculation results clearly indicate that the parallel flow $\mathbf{k}\cdot\mathbf{u}_0$  on the $q=2$ resonant surface gradually slows down and eventually drops to zero as a consequence of the resonant electromagnetic torque induced by the RMP, whereas the parallel flow in the core region slows down as well but remains finite.

\begin{figure}[htbp]
\centering
\subfigure{\includegraphics[width=0.49\textwidth,angle=0]{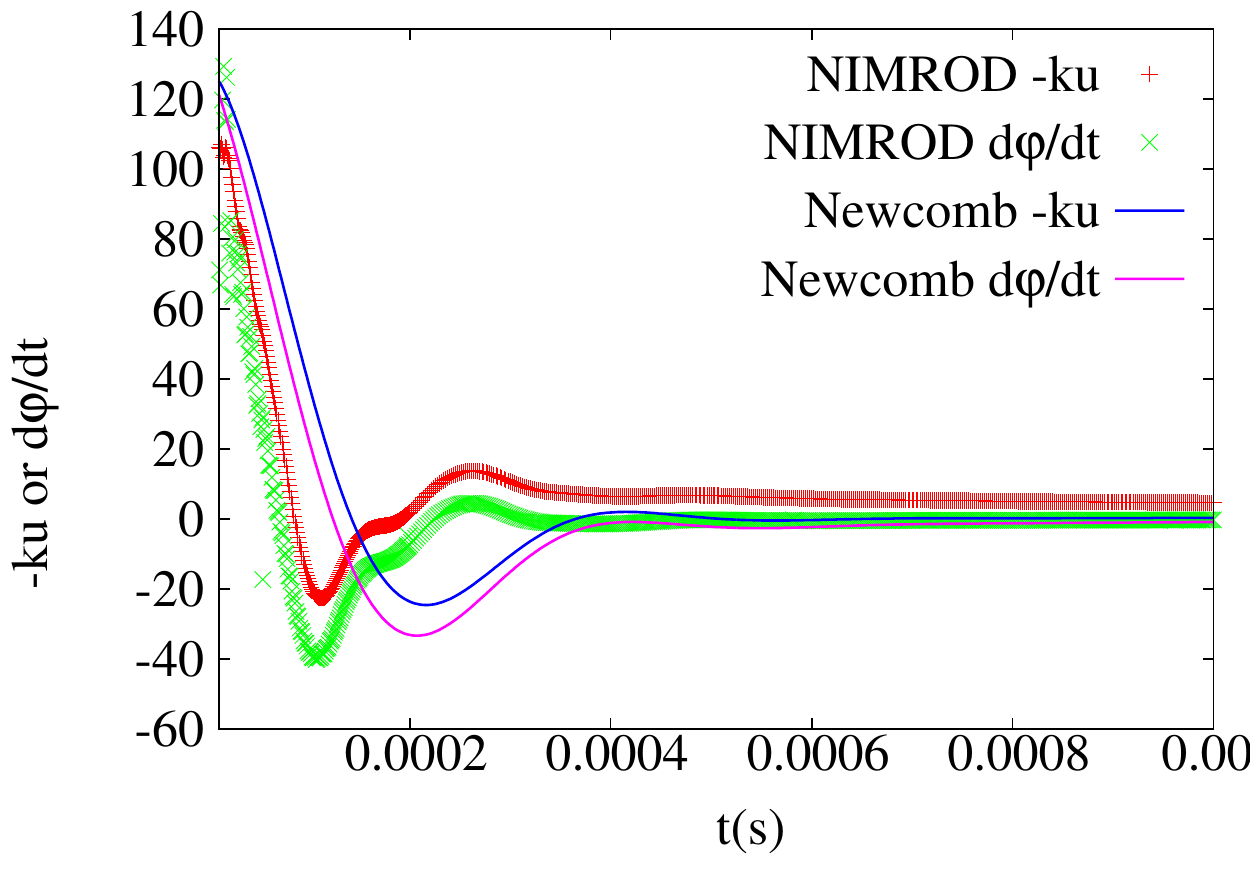}}
\subfigure{\includegraphics[width=0.49\textwidth,angle=0]{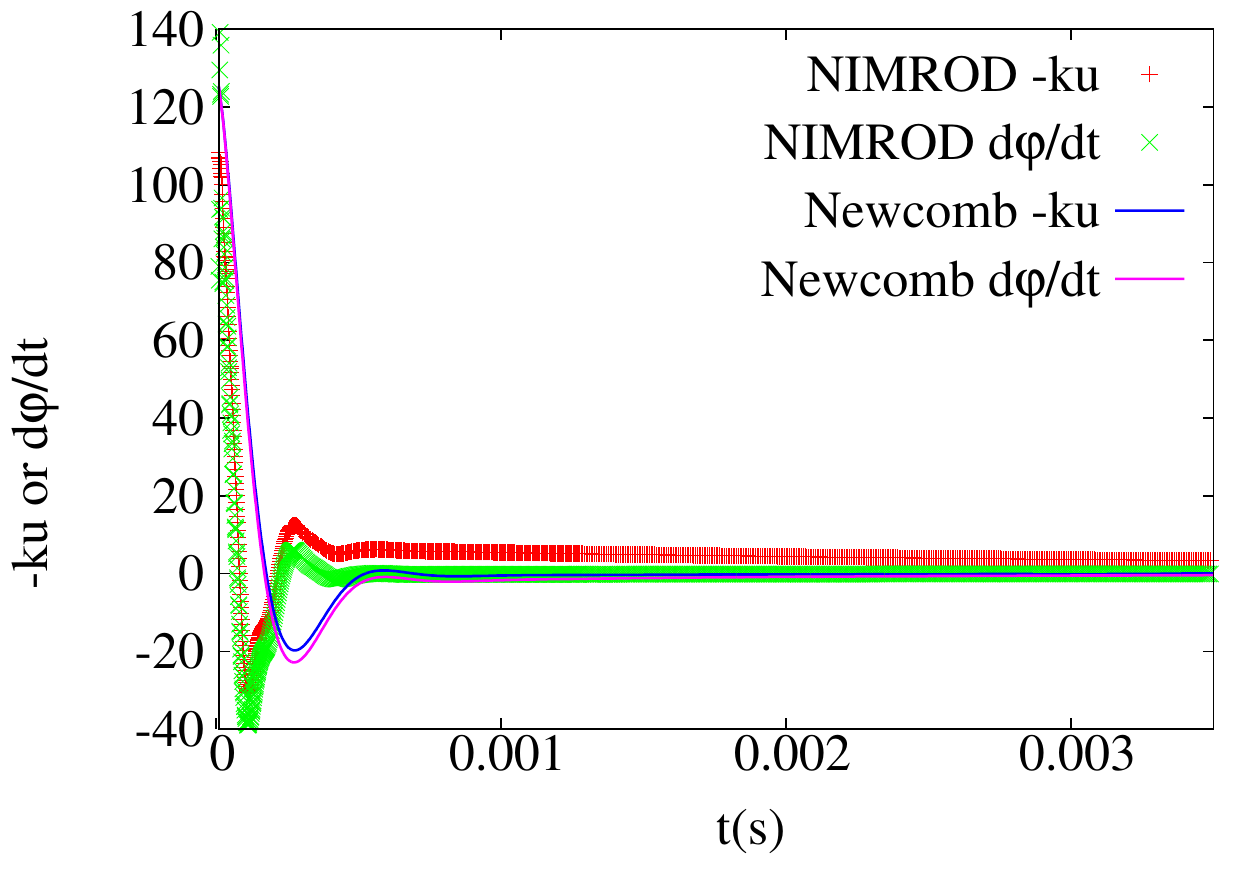}}

\subfigure{\includegraphics[width=0.49\textwidth,angle=0]{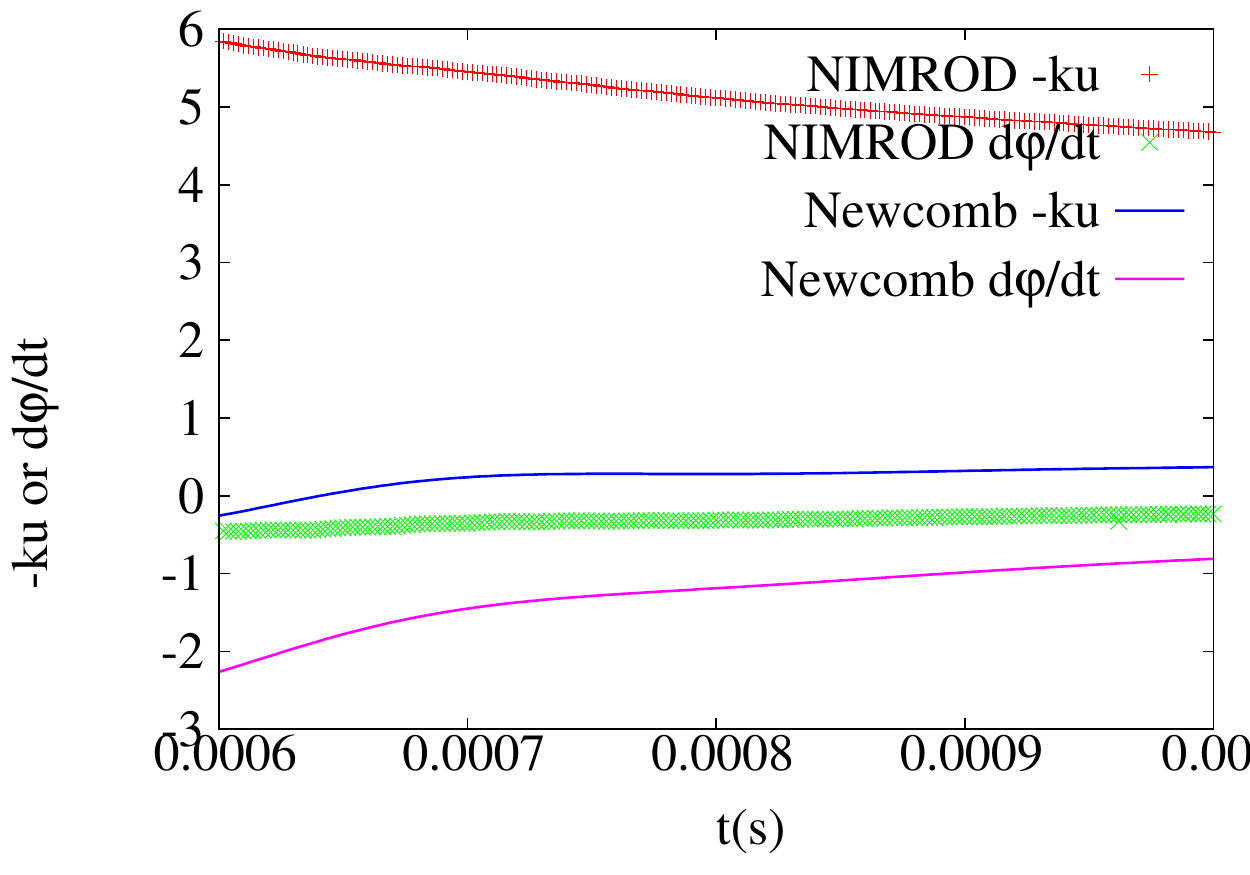}}
\subfigure{\includegraphics[width=0.49\textwidth,angle=0]{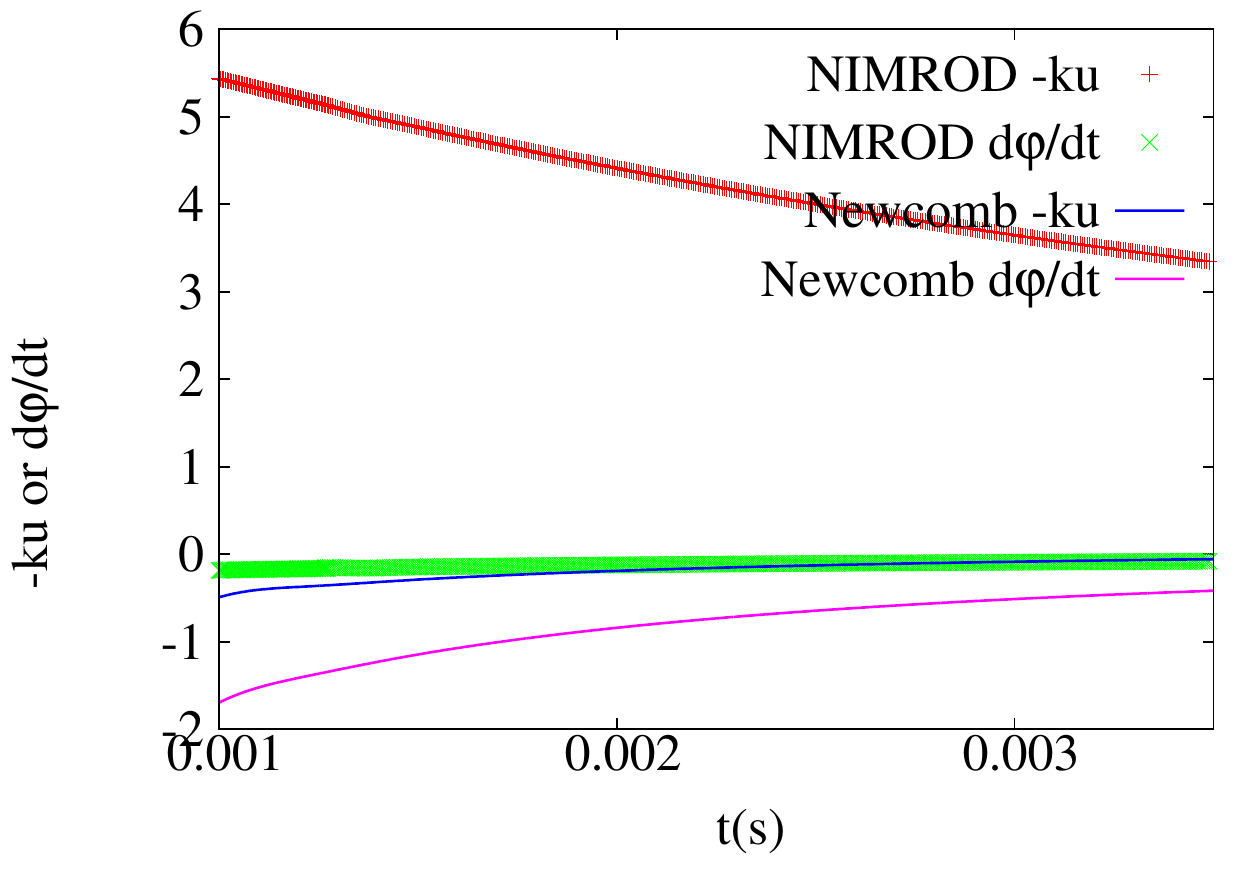}} 
\caption{Phase change rates of plasma response $d\varphi/dt$ and flow $\mathbf{k}\cdot\mathbf{u}_0$ at $q=2$ surface as functions of time over entire course (upper) and approaching steady state (lower) for the $S=3\times10^5$, $Pr_m=40$ (left) and the $S=3\times10^6$, $Pr_m=400$ cases (right).}
\label{fig:locked_ku_history}
\end{figure}

The predictions on $\mathbf{k}\cdot\mathbf{u}_0$ profile evolution from NS and FS theory models are almost identical for both the``locked state'' cases without obvious difference as can be seen from Fig.~\ref{fig:locked_ku_profile}. The difference in the two theory models in terms of the phase relation between plasma flow and resonant response, however, does show up in Fig.~\ref{fig:locked_ku_history}, where the $\mathbf{k}\cdot\mathbf{u}_0$ and the $d\varphi/dt$ on the resonant surface from NIMROD simulations and Newcomb numerical solutions to the FS theory model are compared entirely and in details as functions of time. For both the $S=3\times10^5$, $Pr_m=40$ case (Fig.~\ref{fig:locked_ku_history}, left column) and the $S=3\times10^6$, $Pr_m=400$ case (Fig.~\ref{fig:locked_ku_history}, right column), there are small yet finite differences between the $\mathbf{k}\cdot\mathbf{u}_0$ and the $d\varphi/dt$ on the $q=2$ surface in the Newcomb solutions to the FS theory model throughout the course, which would be exactly absent in the Newcomb solutions to the NS theory model. Such finite differences between the $\mathbf{k}\cdot\mathbf{u}_0$ and the $d\varphi/dt$ in the Newcomb solutions are an approximation to the corresponding differences within the NIMROD simulations, which are more significant in both ``locked state'' cases (Fig.~\ref{fig:locked_ku_history}). Apparently, the ``no-slip'' condition assumed in the NS theory model is less applicable here. Also note that the values of $d\varphi/dt$ are less than those of $\mathbf{k}\cdot\mathbf{u}_0$ in the Newcomb solutions throughout the time, which is consistent with both the phase equation~(\ref{eq:phase_fslip}) in the FS theory model and the NIMROD simulation results (Fig.~\ref{fig:locked_ku_history}).  As explained earlier in Sec.~\ref{sec:the}, the difference in phase-change between the plasma response and flow on the resonant surface, i.e. the ``slipping'', is due to the finite resistivity and the consequential break-down of frozen-in condition there. Such a difference between the $\mathbf{k}\cdot\mathbf{u}_0$ and the $d\varphi/dt$ should vanish as the resistivity approaches zero or $S\to\infty$, in the more collisionless plasma regime. This is indeed the case in both NIMROD simulations and the Newcomb solutions to the FS theory model, as can be seen from comparing the two ``locked state'' cases with different Lundquist and magnetic Prandtl numbers shown in the two columns of Fig.~\ref{fig:locked_ku_history}.

\subsection{Unlocked state of plasma flow}
\label{subsec:unlocked}
\begin{figure}[htbp]
\centering
\subfigure{\includegraphics[width=0.49\textwidth,angle=0]{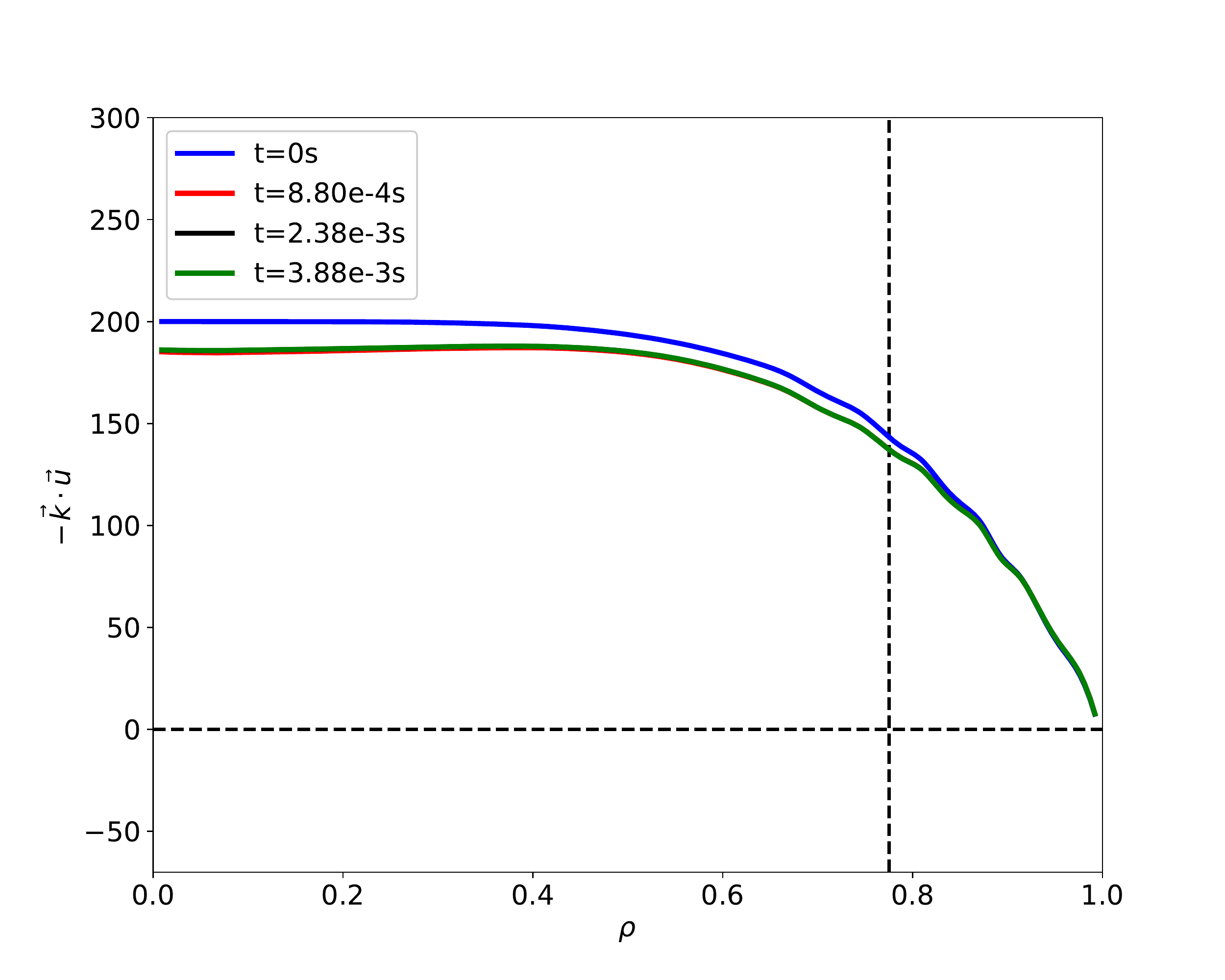}}
\subfigure{\includegraphics[width=0.49\textwidth,angle=0]{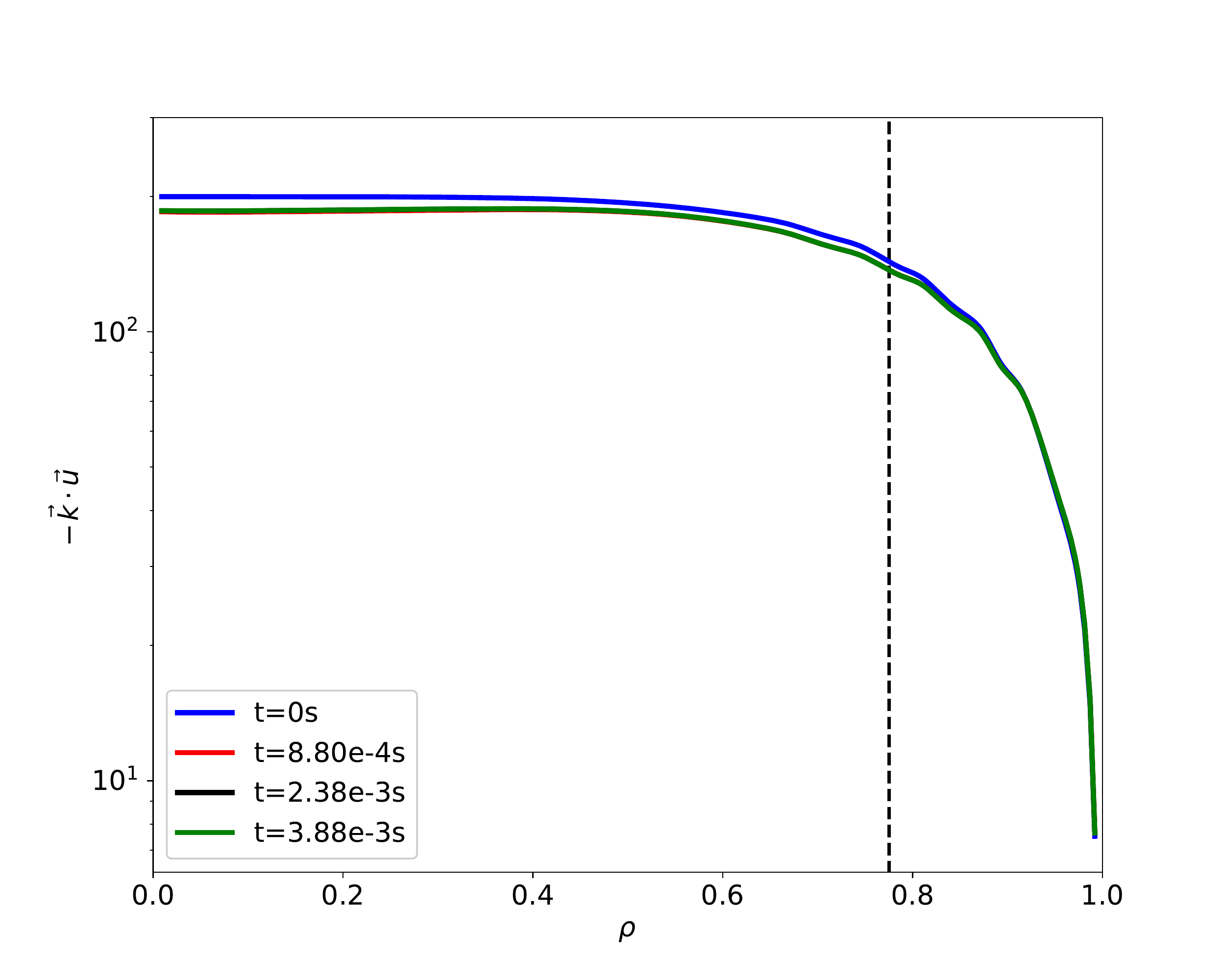}}

\subfigure{\includegraphics[width=0.49\textwidth,angle=0]{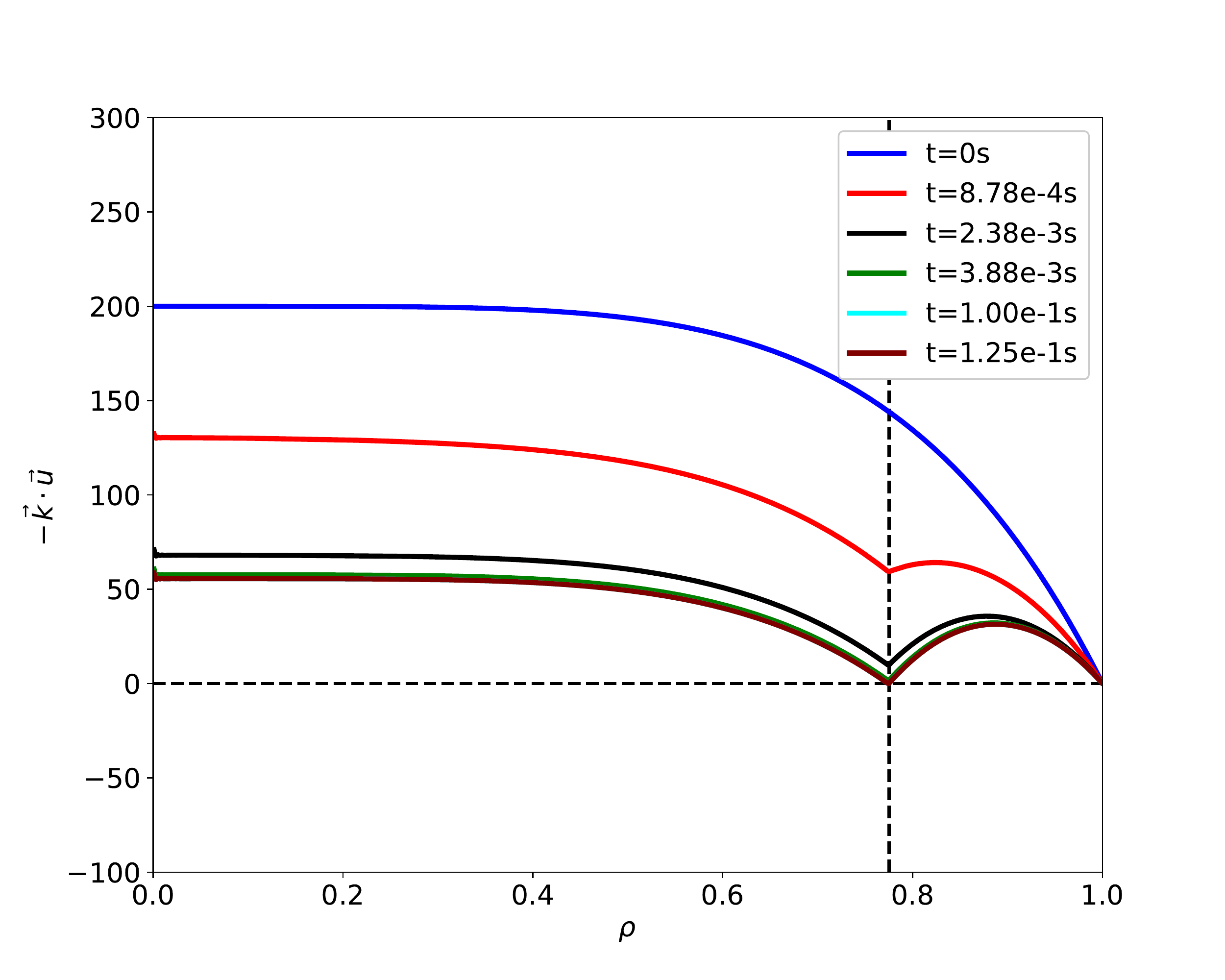}}
\subfigure{\includegraphics[width=0.49\textwidth,angle=0]{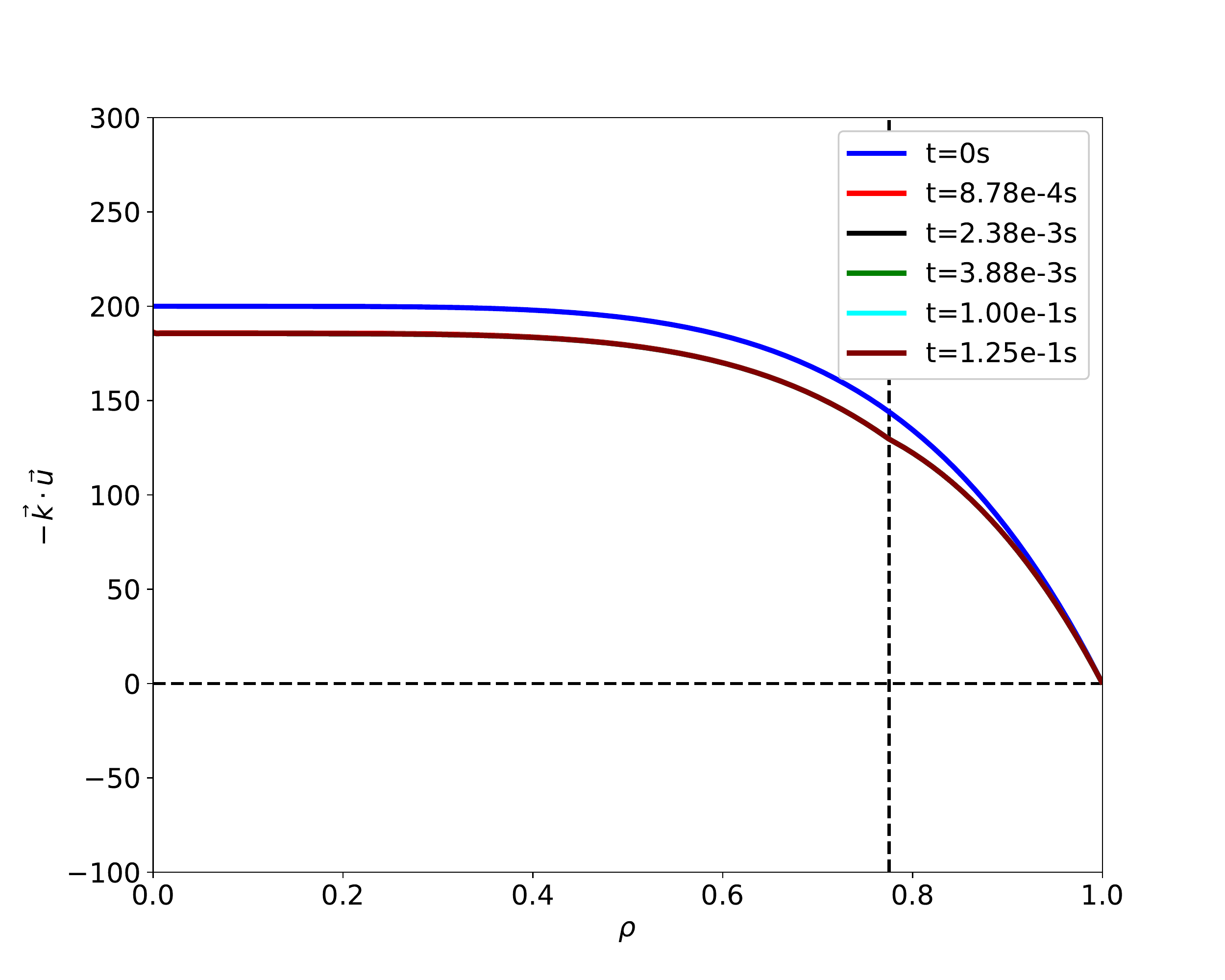}}

\subfigure{\includegraphics[width=0.49\textwidth,angle=0]{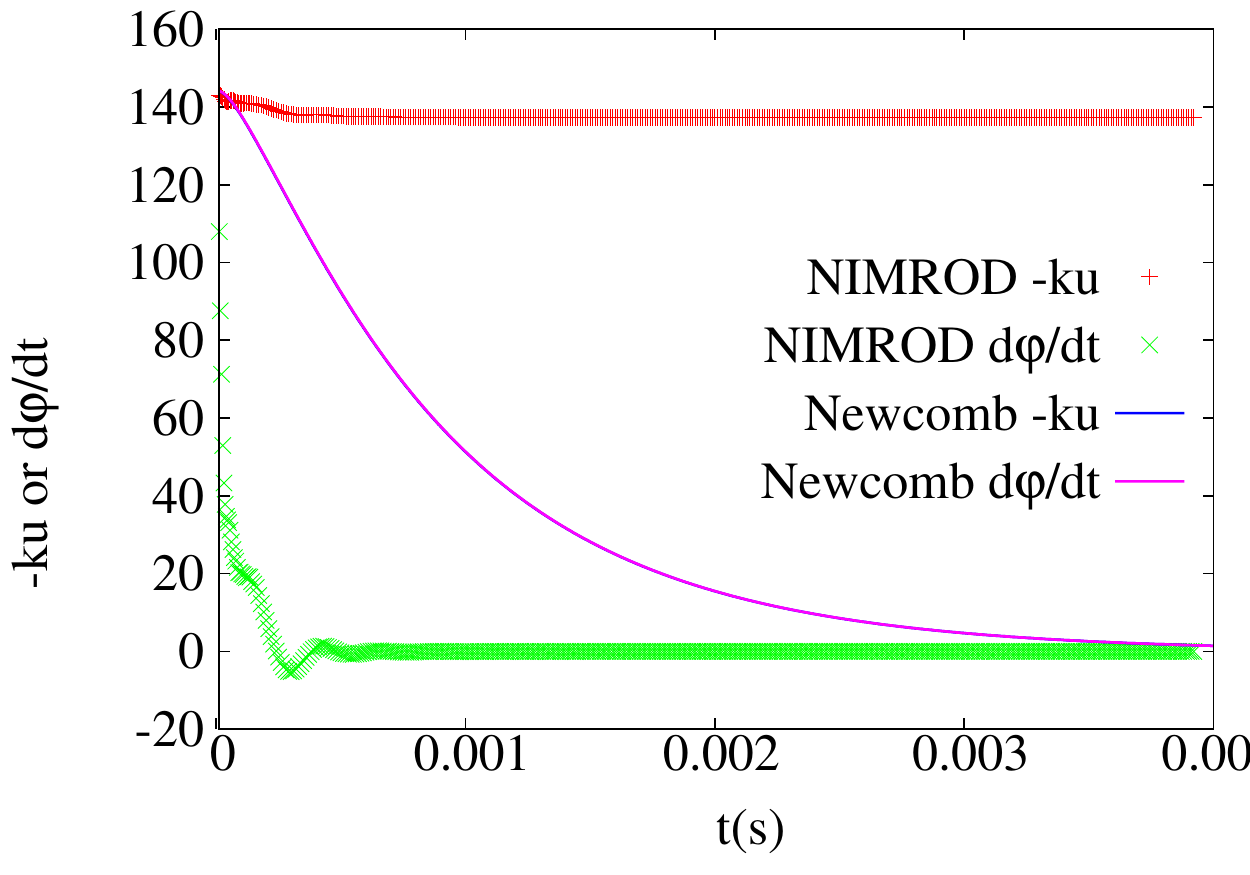}}  
\subfigure{\includegraphics[width=0.49\textwidth,angle=0]{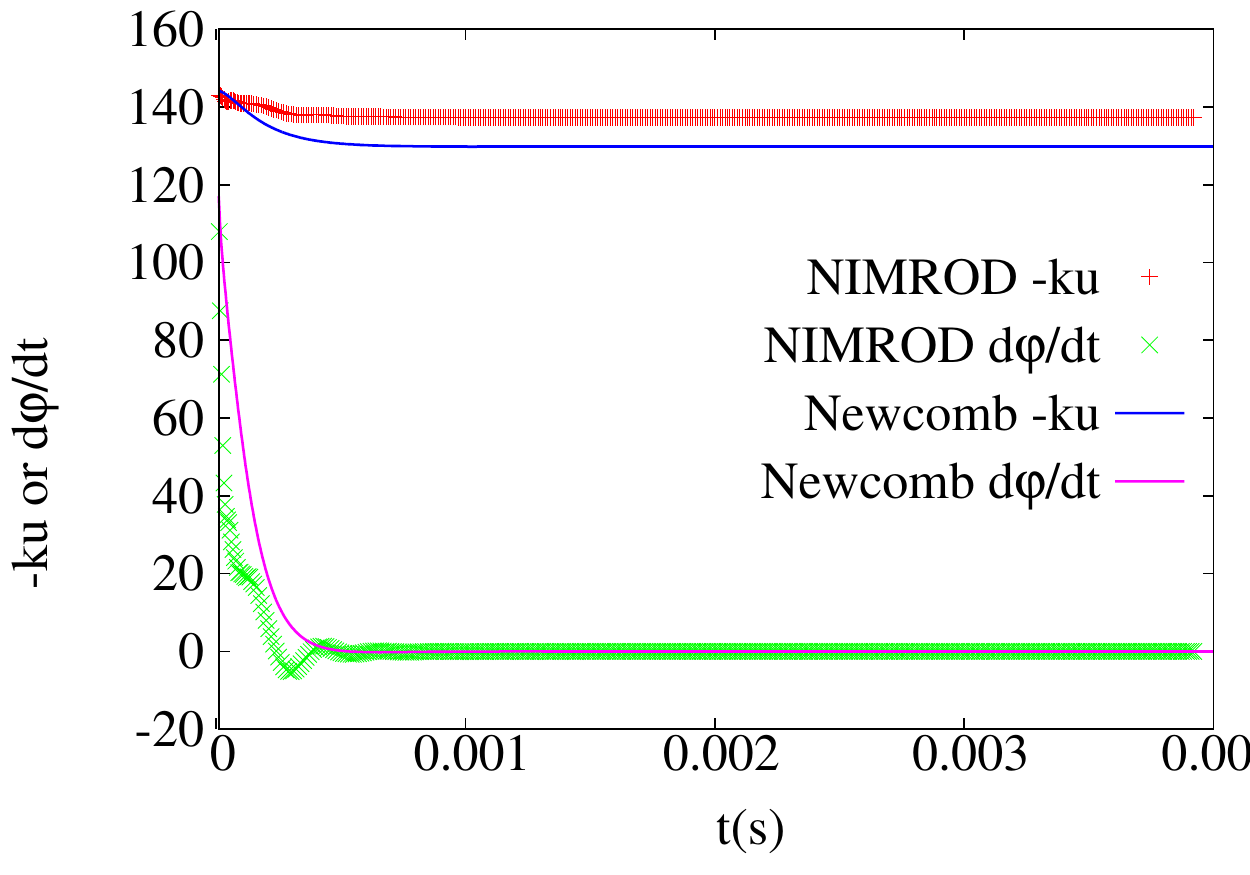}}  
\caption{Radial profiles of $\mathbf{k}\cdot\mathbf{u}_0$ at different time slices from NIMROD simulations (in linear scale for vertical axis, upper left; in logarithmic scale for vertical axis, upper right), and Newcomb solutions to the NS theory model (middle left) and the FS theory model (middle right), and phase change rates of plasma response $d\varphi/dt$ and flow $\mathbf{k}\cdot\mathbf{u}_0$ at $q=2$ surface as functions of time from NIMROD simulations and Newcomb solutions to the NS theory model (lower left) and the FS theory model (lower right). Here $S=2.44\times10^3$, $Pr_m=1$, and $\Omega_0=2\times 10^2rad/s$.}
\label{fig:unlocked_1_ku_profile_history}
\end{figure}

For a sufficiently large rotation magnitude, weak RMP amplitude, or strong viscosity, nonlinear plasma response to RMP can reach a steady state where the plasma flow remains finite on the resonant flux surface, which is referred to as the ``unlocked state'' of plasma flow. Here we report two cases of such unlocked states in response to RMP, where the uniform number density $N=10^{19}m^{-3}$, and the equivalent island width of the RMP amplitude $W_C/a=0.146$. The first case is in a more resistive regime, with $S=2.44\times10^3$, $Pr_m=1$, and $\Omega_0=2\times 10^2rad/s$.  For this case, however, Newcomb solutions to the NS theory model predict a ``locked state'' of plasma flow where $\mathbf{k}\cdot\mathbf{u}_0$ drops to zero in the final steady state of plasma response (Fig.~\ref{fig:unlocked_1_ku_profile_history}, middle left). In contrast, Newcomb solutions to the FS theory model predict a ``unlocked state'' of plasma flow where $\mathbf{k}\cdot\mathbf{u}_0$ remains finite in the final steady state of plasma response (Fig.~\ref{fig:unlocked_1_ku_profile_history}, middle left), which agrees with the results from NIMROD simulations (Fig.~\ref{fig:unlocked_1_ku_profile_history}, upper panels). The ``no-slip'' condition simply does not apply in the regime represented by this case. 
The phase change rates of plasma response $d\varphi/dt$ and flow $\mathbf{k}\cdot\mathbf{u}_0$ at $q=2$ surface as functions of time from NIMROD simulations also agree better with Newcomb solutions to the FS theory model (Fig.~\ref{fig:unlocked_1_ku_profile_history}, lower right) than to the NS theory model (Fig.~\ref{fig:unlocked_1_ku_profile_history}, lower left), where in the NS model the difference between $d\varphi/dt$ and $\mathbf{k}\cdot\mathbf{u}_0$ is assumed zero. Both $d\varphi/dt$ and $\mathbf{k}\cdot\mathbf{u}_0$ from Newcomb solutions to the FS theory model agree well with the NIMROD simulation results, including their relative magnitudes and difference, similar to the cases of ``locked states'' of plasma flow in Sec.~\ref{subsec:unlocked}.

\begin{figure}[htbp]
\centering
\subfigure{\includegraphics[width=0.49\textwidth,angle=0]{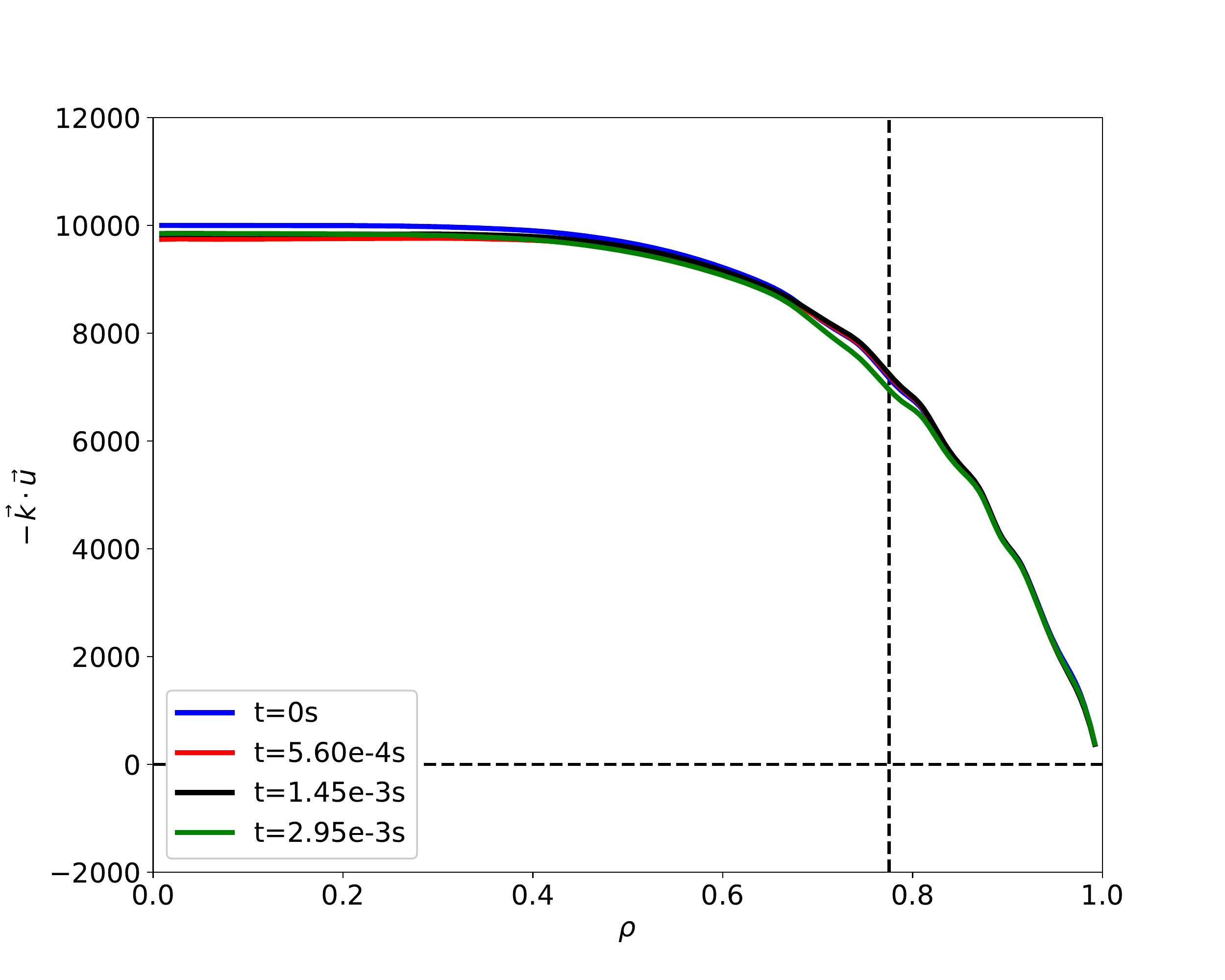}}
\subfigure{\includegraphics[width=0.49\textwidth,angle=0]{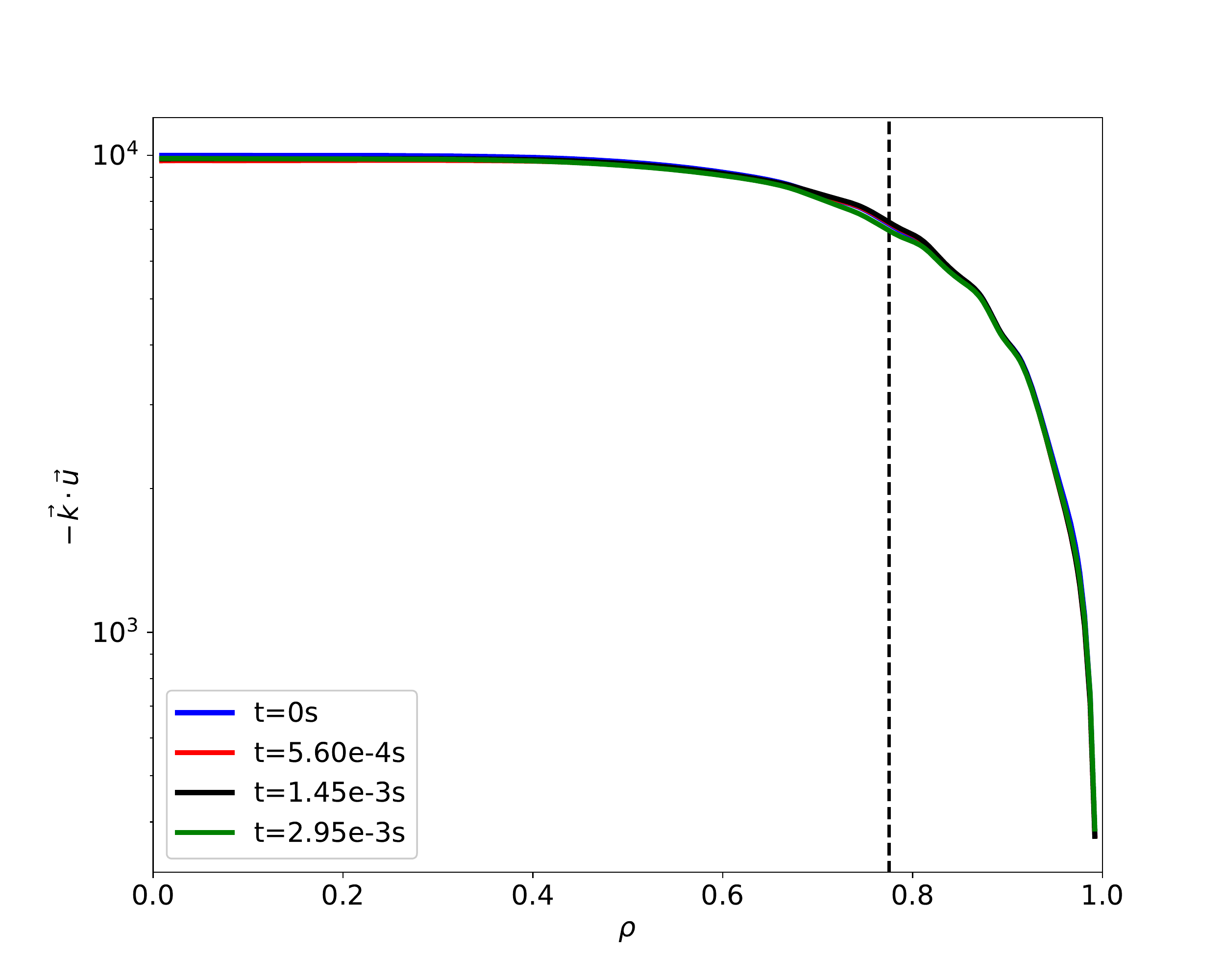}}

\subfigure{\includegraphics[width=0.49\textwidth,angle=0]{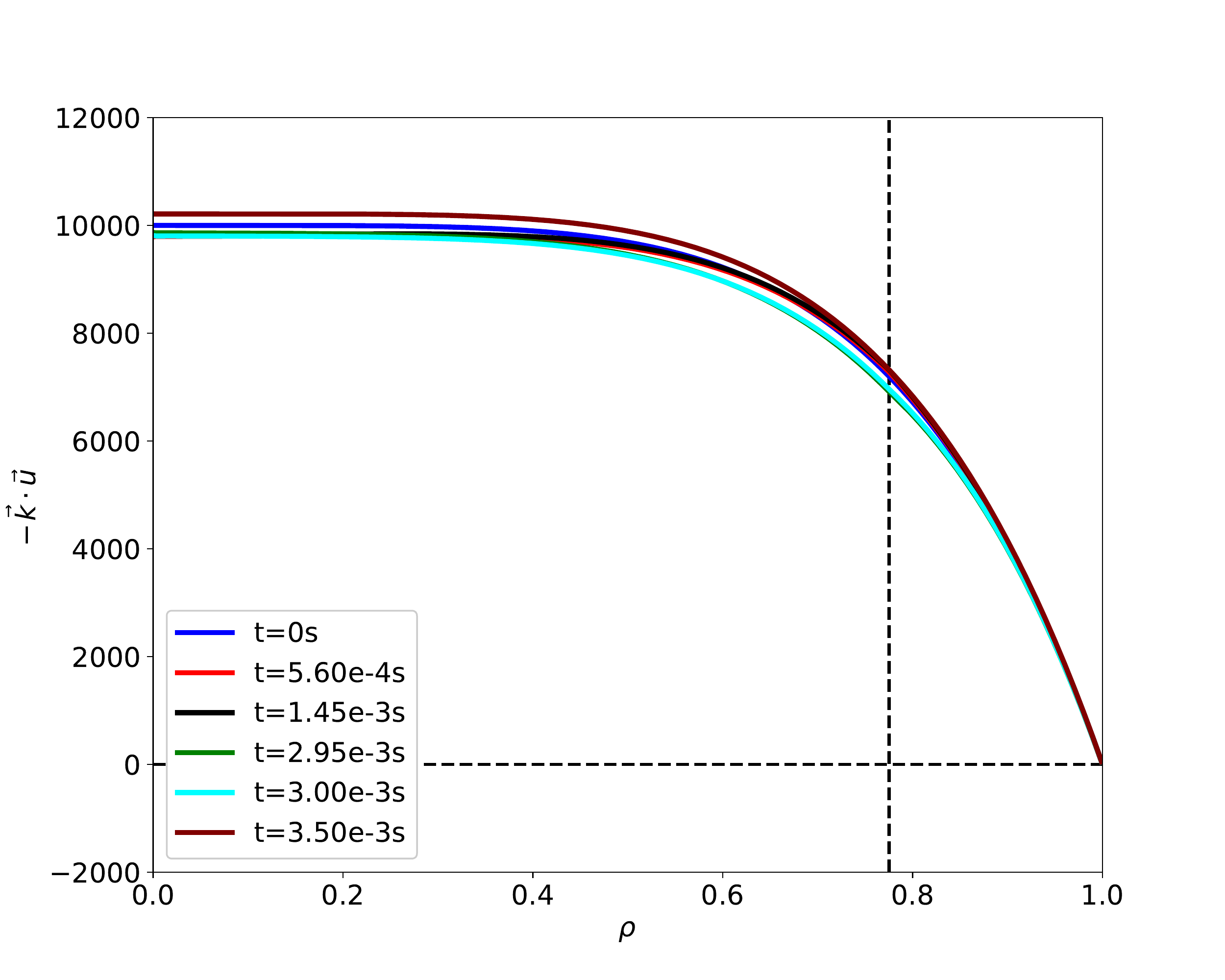}}
\subfigure{\includegraphics[width=0.49\textwidth,angle=0]{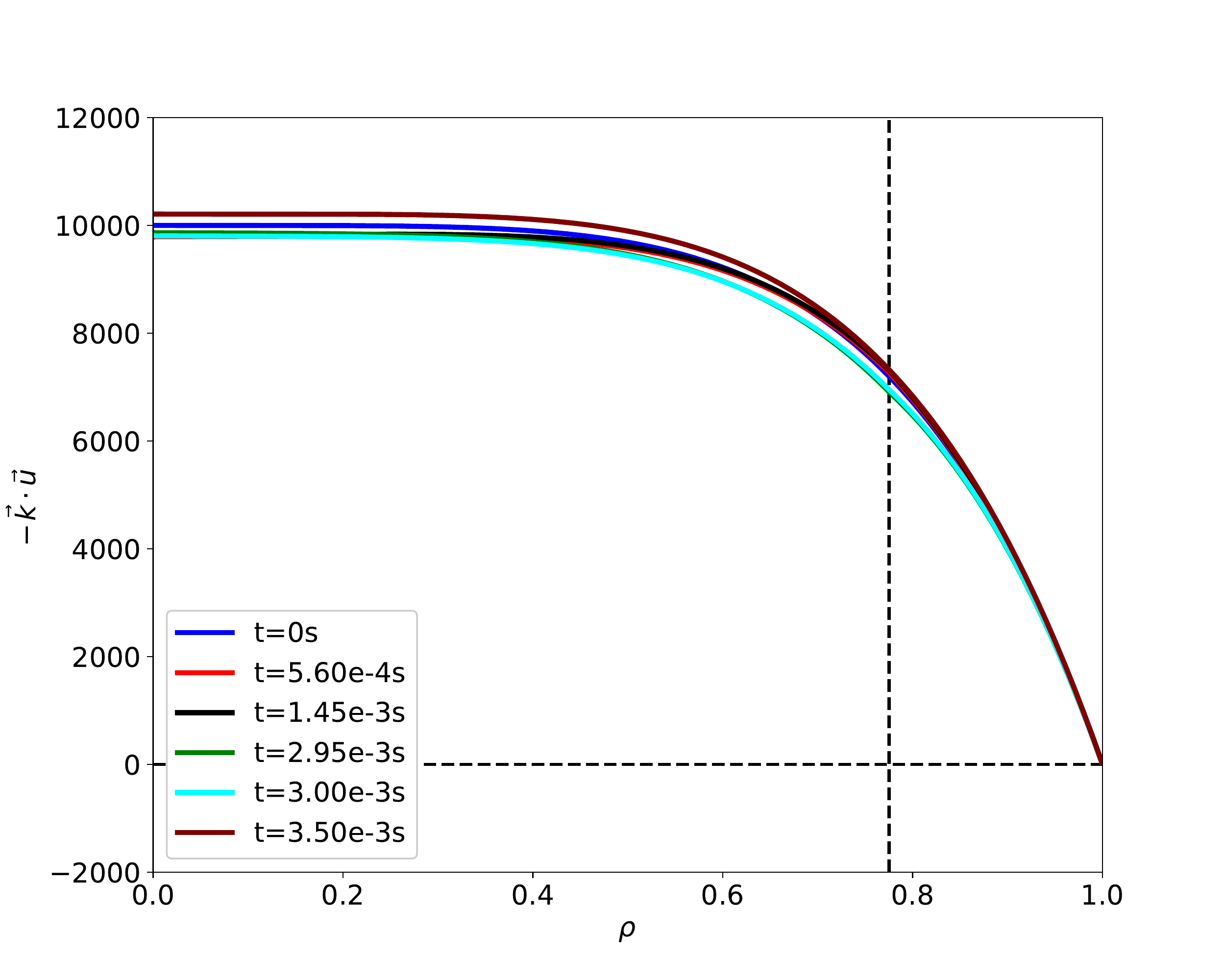}}

\subfigure{\includegraphics[width=0.49\textwidth,angle=0]{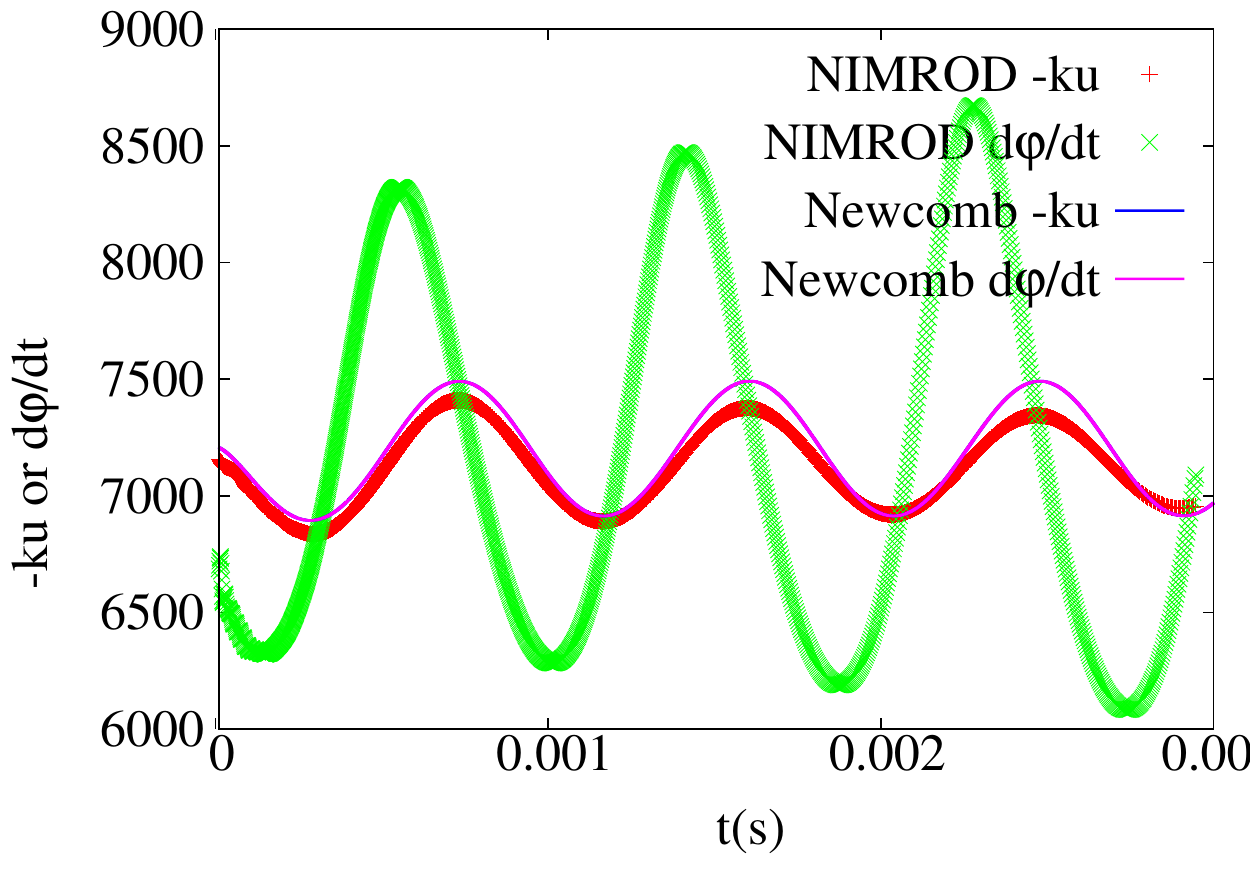}}  
\subfigure{\includegraphics[width=0.49\textwidth,angle=0]{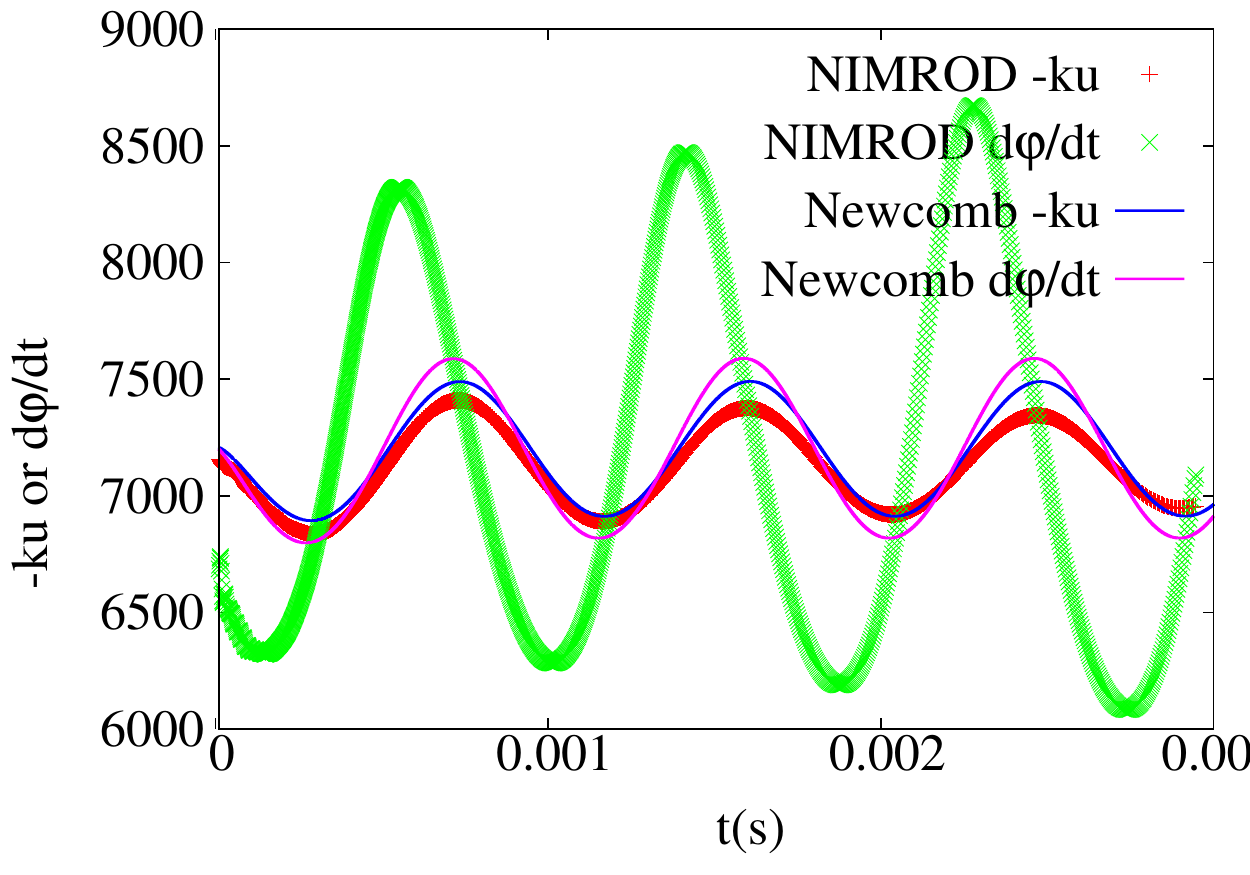}}
\caption{Radial profiles of $\mathbf{k}\cdot\mathbf{u}_0$ at different time slices from NIMROD simulations (in linear scale for vertical axis, upper left; in logarithmic scale for vertical axis, upper right), and Newcomb solutions to the NS theory model (middle left) and the FS theory model (middle right), and phase change rates of plasma response $d\varphi/dt$ and flow $\mathbf{k}\cdot\mathbf{u}_0$ at $q=2$ surface as functions of time from NIMROD simulations and Newcomb solutions to the NS theory model (lower left) and the FS theory model (lower right). Here $S=10^6$, $Pr_m=400$, and $\Omega_0=10^4rad/s$.}
\label{fig:unlocked_2_ku_profile_history}
\end{figure}

The second case of ``unlocked state'' is in a less resistive regime where $S=10^6$, $Pr_m=400$, and $\Omega_0=10^4rad/s$. In this case, the Newcomb solutions to both NS and FS theory models predict an ``unlocked state'' with parallel plasma flow $\mathbf{k}\cdot\mathbf{u}_0$ remaining finite on the $q=2$ surface. The $\mathbf{k}\cdot\mathbf{u}_0$ profiles from both theory models agree with each other (Fig.~\ref{fig:unlocked_2_ku_profile_history}, middle panels) and with the NIMROD simulation results (Fig.~\ref{fig:unlocked_2_ku_profile_history}, upper panels). This is understandable, since the difference between the NS and FS theory models diminishes in the collisionless or ideal regime, where the ``no-slip'' condition should be more relevant. Despite the good agreement among theory model solutions and the NIMROD simulation results in terms of the parallel flow $\mathbf{k}\cdot\mathbf{u}_0$ profile evolution, there are still some finite differences between the phase change rates of plasma response $d\varphi/dt$ and flow $\mathbf{k}\cdot\mathbf{u}_0$ at the $q=2$ surface, which is rather small from the FS theory model prediction, but remains quite large in the NIMROD simulation results (Fig.~\ref{fig:unlocked_2_ku_profile_history}, lower panels). Unlike all other cases of both ``locked states'' and ``unlocked states'' of plasma flow presented earlier, the initial toroidal rotation frequency is $100$ times larger in this case, which contributes to the appearance of several oscillating periods within the same time frame of evolution for $d\varphi/dt$ and $\mathbf{k}\cdot\mathbf{u}_0$ at the $q=2$ surface in both FS theory solutions and NIMROD simulations here, which is also consistent with the phase equation in Eq.~(\ref{eq:phase_fslip}). In addition, although the time history of $\mathbf{k}\cdot\mathbf{u}_0$ at the $q=2$ surface from the FS theory solution agrees well with that from the NIMROD simulation, the oscillation of $d\varphi/dt$ from the FS theory solution does not track that from the NIMROD simulation well in terms of either amplitude or phase (Fig.~\ref{fig:unlocked_2_ku_profile_history}, lower right). The fact that this discrepancy shows up so far only in the case of larger initial toroidal flow, may suggest further room for improvement on the modeling of the toroidal rotation dynamics in the theory of plasma response. 

\section{Summary and discussion}
\label{sec:sum}
In summary, the plasma flow evolution in response to RMP in a tokamak has been evaluated within the resistive single-fluid MHD model using analytical theories and NIMROD simulations. Representative cases for both ``locked'' and ``unlocked'' states of the parallel plasma flow along with the steady states of nonlinear response are considered and reported. Good agreement between NIMROD simulations and numerical solutions to the extended theory with the ``free-slip'' condition has been achieved for the parallel flow profile evolution in response to RMP in all resistive regimes, whereas the difference from the conventional theory with the ``no-slip'' condition tends to diminish as the plasma resistivity approaches zero. As predicted from theory, the``no-slip'' condition for the phase relation between the nonlinear plasma response and the parallel plasma flow becomes more applicable in the high Lundquist number $S$ regime. However, even the extended theory allowing ``free-slip'' condition is unable to capture the remaining and substantial difference in the phase change rate between the plasma response and the parallel plasma flow on the resonant flux surface obtained from NIMROD simulations in the high Lundquist number $S$ regime with larger initial equilibrium toroidal rotation. This suggests that the theory of nonlinear plasma response needs further improvement on the part of toroidal rotation dynamics.

Beyond the resistive single-fluid MHD model, two-fluid and kinetic effects, 2D and 3D neoclassical effects including those from neoclassical toroidal viscosity (NTV) induced from non-resonant response, are all necessary to adequately and self-consistently account for the realistic physics, including the plasma flow response involved in the RMP experiments. Although much efforts have been devoted to the study of these effects, their individual roles, relative importance, and integrated significance remain subjects of intensive research in the near future.


\acknowledgments
This work was supported by the Fundamental Research Funds for the Central Universities at Huazhong University of Science and Technology Grant No. 2019kfyXJJS193, the National Natural Science Foundation of China Grant Nos. 11775221 and 51821005, the Young Elite Scientists Sponsorship Program by CAST Grant No. 2017QNRC001, and the U.S. Department of Energy Grant Nos. DE-FG02-86ER53218 and DE-SC0018001. The authors are grateful for the helpful discussions with Profs. C.~C. Hegna and C.~R. Sovinec, the assistance by Fangyuan Ma, and the support from the NIMROD team. The data that support the findings of this study are available from the corresponding author upon reasonable request.

\appendix
\section{Equations for toroidal and poloidal rotations in a cylindrical tokamak in presence of RMP}
\label{app:tor_pol}
The toroidal and poloidal flows in a tokamak can be written as following
\begin{align}
&\Omega_\phi=\Omega_{\phi 0}+\Delta\Omega_\phi, \\
&\Omega_\theta=\Omega_{\theta 0}+\Delta\Omega_\theta,
\end{align}
where $\Omega_{\phi 0}$ ($\Delta\Omega_{\phi}$) and $\Omega_{\theta 0}$ ($\Delta\Omega_{\theta}$) are the equilibrium (perturbed) rotation frequencies in the toroidal and poloidal directions, respectively. The equation and initial-boundary conditions for $\Delta\Omega_\phi$ in a cylindrical tokamak are
\begin{align}
&r\rho\frac{\partial \Delta\Omega_\phi}{\partial t}-\frac{\partial}{\partial r}\left(r\mu\frac{\partial \Delta\Omega_\phi}{\partial r}\right)=\frac{\hat{T_z}}{4\pi^2R_0^3}\delta(r-r_s),\\
&\frac{\partial\Delta\Omega_\phi(0,t)}{\partial r}=\Delta\Omega_\phi(a,t)=0,
\end{align}
where the detailed expression for the toroidal electromagnetic torque $\hat{T_z}$ can be found in Ref.~\cite{huangwl15a}. Using Bessel function expansion following Refs.~\cite{zanca08a,licg14a}, we have
\begin{align}
\Delta\Omega_\phi(r,t)=\sum_{j=1}^{\infty}a_jJ_0(j_{0,j}\frac{r}{a}),
\end{align}
then the toroidal rotation equation in Bessel spectral space becomes
\begin{align}
a^2\rho\left(\frac{\partial a_j}{\partial t}+a_j\frac{j_{0,j}^2}{\tau_V}\right)\frac{1}{2}J_1^2(j_{0,j})=\frac{\hat{T_z}}{4\pi^2R_0^3}J_0(j_{0,j}\frac{r_s}{a})=-\frac{1}{2}C_1W^2W_c^2\sin\varphi J_0(j_{0,j}\frac{r_s}{a}),
\end{align}
where $\tau_V=\frac{a^2\rho}{\mu}$, and $j_{0,j}$ are the zero points of the zeroth order Bessel function $J_0$.

The equations and initial-boundary condition for $\Delta\Omega_\theta$ in a cylindrical tokamak are
\begin{align}
&r^3\rho\left(\frac{\partial\Delta\Omega_\theta}{\partial t}+\frac{\Delta\Omega_\theta}{\tau_D}\right)-\frac{\partial}{\partial r}\left(r^3\mu\frac{\partial\Delta\Omega_\theta}{\partial r}\right)=-\frac{m\hat{T_z}}{n4\pi^2R_0}\delta(r-r_s),\\
&\frac{\partial\Delta\Omega_\theta(0,t)}{\partial r}=\Delta\Omega_\theta(a,t)=0,
\end{align}
where $\tau_D$ is a model damping time for the poloidal rotation. Similarly, following Refs.~\cite{zanca08a,licg14a}, we expand $\Delta\Omega_\theta$ as following
\begin{align}
\Delta\Omega_\theta(r,t)=\sum_{j=1}^\infty f_j(t)\nu_j(r).
\end{align}
Then, we have
\begin{align}
\rho\left(\frac{\partial f_j}{\partial t} + \frac{j_{1,j}^2}{\tau_V} f_j + \frac{f_j}{\tau_D}\right) = -\frac{m\hat{T_z}}{n4\pi^2R_0}\nu_j(r_s),
\end{align}
where $\nu_j$ satisfies 
\begin{align}
&\frac{d}{dr}\left(r^3\mu\frac{d\nu_j}{dr}\right)+r^3\rho\gamma_j\nu_j=0,\\
&\frac{d\nu_j(0)}{dr}=\nu_j(a)=0.
\end{align}
Here, $j_{1,j}$ are the zero points of the first order Bessel function $J_1$, and
\begin{align}
&\nu_j(r)=E_j\frac{J_1(j_{1,j}\frac{r}{a})}{r}, \gamma_j=\frac{j_{1,j}^2}{\tau_V}, \\ &E_j=\left[\int_0^arJ_1^2(j_{1,j}\frac{r}{a})dr\right]^{-1/2}=\left[\frac{a^2}{2}J_2^2(j_{1,j})\right]^{-1/2}.
\end{align}
The above equations and their expansions in the Bessel functional space are solved numerically to obtain the ``Newcomb'' solutions to the theory models for nonlinear plasma response adopted in this work.

\section{Analytical solution for the steady state nonlinear plasma response to RMP in presence of both toroidal and poloidal rotations}
\label{app:w_wc}
Assuming $\tau_D\to\infty$, the steady state solutions for $\Omega_\theta$ and $\Omega_\phi$ are
\begin{align}
\Omega_\theta&=\Omega_{\theta 0}+\Delta\Omega_\theta \\
&=\Omega_{\theta 0}+\frac{mC_1R_0^2W^2W_c^2\sin\varphi}{2n\rho}\sum_j\frac{\nu_j(r_s)\nu_j(r)}{j_{1,j}^2/\tau_V},
\end{align}
and
\begin{align}
\Omega_\phi&=\Omega_{\phi 0}+\Delta\Omega_\phi \\
&=\Omega_{\phi 0}-\sum_j\frac{C_1W^2W_c^2\sin\varphi}{\mu j_{0,j}^2J_1^2(j_{0,j})}J_0(j_{0,j}\frac{r_s}{a})J_0(j_{0,j}\frac{r}{a}).
\end{align}

At the rational surface $r=r_s$, the perturbed toroidal flow can be expressed as
\begin{align}
\Omega_{\phi s}=\Omega_{\phi 0}-\frac{C_1W^2W_c^2\sin\varphi}{\mu}\sum_j\frac{J_0^2(j_{0,j}\frac{r_s}{a})}{j_{0,j}^2J_1^2(j_{0,j})}.
\end{align}
To be simple, we define
\begin{align}
C_2=\sum_j\frac{J_0^2(j_{0,j}\frac{r_s}{a})}{j_{0,j}^2J_1^2(j_{0,j})}.
\end{align}
Then,
\begin{align}
\Omega_{\phi s}=\Omega_{\phi 0}-\frac{C_1C_2W^2W_c^2\sin\varphi}{\mu}
\end{align}
Similarly, the poloidal flow at the rational surface can be expressed as following
\begin{align}
\Omega_{\theta s}=\Omega_{\theta 0}+\frac{mC_1C_3R_0^2W^2W_c^2\sin\varphi}{2n\rho},
\end{align}
where
\begin{align}
C_3=\sum_j\frac{\nu_j^2(r_s)}{j_{1,j}^2/\tau_V}.
\end{align}

Combining the no-slip condition and the island width evolution equation, we obtain the relation between the steady state nonlinear plasma response amplitude $W$ and the RMP amplitude $W_c$ from the NS theory model as following~\cite{huangwl15a}
\begin{align}
W_c=\left[\frac{(n\Omega_{\phi 0}-m\Omega_{\theta 0})^2}{W^4\left(\frac{\displaystyle nC_1C_2}{\displaystyle\mu}+\frac{\displaystyle m^2R_0^2C_1C_3}{\displaystyle 2n\rho}\right)^2}+W^4\frac{(\Delta_l'+\Delta_{nl}')^2}{{\Delta_c'}^2}\right]^{1/4}.
\end{align}

Similarly, combining the steady state free-slip condition
\begin{align}
n\Omega_{\phi s}-m\Omega_{\theta s}=\frac{\sqrt2a^2}{2A\tau_R}\Delta_c'\frac{W_c^2}{W^3}\sin\varphi,
\end{align}
and island width evolution equation,
we obtain the relation between $W$ and $W_c$ for the steady state of nonlinear plasma response from the FS theory model as following
\begin{align}
W_c=\left[\frac{(n\Omega_{\phi 0}-m\Omega_{\theta 0})^2}{W^4\left(\frac{\displaystyle nC_1C_2}{\displaystyle\mu}+\frac{\displaystyle m^2R_0^2C_1C_3}{\displaystyle 2n\rho}\right)^2 + \frac{\displaystyle\left(\frac{\sqrt2a^2}{2A\tau_R}\Delta_c'\right)^2}{\displaystyle W^6}}+W^4\frac{(\Delta_l'+\Delta_{nl}')^2}{{\Delta_c'}^2}\right]^{1/4}.
\end{align}

\newpage

\newpage

\newpage



\newpage




\newpage

\newpage

\end{document}